\documentclass{aa}
\input{psfig.sty}
\usepackage{lscape}
\usepackage{graphicx,times}
\usepackage{float}
\usepackage{longtable}

\begin{document}
\authorrunning{K. C. Steenbrugge et al.}
\titlerunning{Physical conditions in the X-ray absorber}
\title{Simultaneous X-ray and UV spectroscopy of the Seyfert galaxy NGC~5548. II. Physical conditions in the X-ray absorber.}
\subtitle{}

\author{K.C. Steenbrugge\inst{1}, J.S. Kaastra\inst{1}, D. M. Crenshaw\inst{2}, S. B. Kraemer\inst{3,4} N. Arav\inst{5}, I. M. George\inst{6,7}, D.~A.~Liedahl\inst{8}, R.~L.~J.~van der Meer\inst{1}, F. B. S. Paerels\inst{9}, T. J. Turner\inst{6,7} and T. Yaqoob\inst{7,10}}
\offprints{K.C. Steenbrugge}
\mail{K.C.Steenbrugge@sron.nl}
\institute{ SRON National Institute for Space Research, Sorbonnelaan 2, 3584 CA Utrecht, The Netherlands
\and Department of Physics and Astronomy, Georgia State University, Astronomy Offices, One Park Place South SE, Suite 700, Atlanta, GA 30303
\and Catholic University of America 
\and Laboratory for Astronomy and Solar Physics, NASA's Goddard Space Flight Center, Code 681, Greenbelt, MD 20771 
\and CASA, University of Colorado, 389 UCB, Boulder, CO 80309-0389, USA
\and Joint Center for Astrophysics, University of Maryland, Baltimore County, 1000 Hilltop Circle, Baltimore, MD 21250
\and Laboratory for High Energy Astrophysics, Code 660, NASA's Goddard Space Flight Center, Greenbelt, MD 20771
\and Physics Department, Lawrence Livermore National Laboratory, PO Box 808, L-41, Livermore, CA 94550
\and Columbia Astrophysics Laboratory, Columbia University, 538W. 120th Street, New York, NY 10027
\and Department of Physics and Astronomy, Johns Hopkins University, Baltimore, MD 21218 }
\date{Received  / Accepted  }

\abstract{We present the results from a 500 ks {\it Chandra} observation of the Seyfert 1 galaxy NGC~5548. We detect broadened (full width half maximum = 8000 km s$^{-1}$) emission lines of \ion{O}{vii} and \ion{C}{vi} in the spectra, similar to those observed in the optical and UV bands. The source was continuously variable, with a 30~\% increase in luminosity in the second half of the observation. The gradual increase in luminosity occurred over a timescale of $\sim$ 300~ks. No variability in the warm absorber was detected between the spectra from the first 170 ks and the second part of the observation. The longer wavelength range of the LETGS resulted in the detection of absorption lines from a broad range of ions, in particular of C, N, O, Ne, Mg, Si, S and Fe. The velocity structure of the X-ray absorber is consistent with the velocity structure measured simultaneously in the ultraviolet spectra. We find that the highest velocity outflow component, at $-$1040~km~s$^{-1}$, becomes increasingly important for higher ionization parameters. This velocity component spans at least three orders of magnitude in ionization parameter, producing both highly ionized X-ray absorption lines (\ion{Mg}{xii}, \ion{Si}{xiv}) as well as UV absorption lines. A similar conclusion is very probable for the other four velocity components. 
\par
Based upon our observations, we argue that the warm absorber probably does not manifest itself in the form of photoionized clumps in pressure equilibrium with a surrounding wind. Instead, a model with a continuous distribution of column density versus ionization parameter gives an excellent fit to our data. From the shape of this distribution and the assumption that the mass loss through the wind should be smaller than the accretion rate onto the black hole, we derive upper limits to the solid angle as small as $10^{-4}$~sr. From this we argue that the outflow occurs in density-stratified streamers. The density stratification across the stream then produces the wide range of ionization parameter observed in this source. We determine an upper limit of 0.3 M$_{\odot}$ yr$^{-1}$ for the mass loss from the galaxy due to the observed outflows.

}

\maketitle

\section{Introduction\label{sect:intro}}
Over half of all Seyfert 1 galaxies exhibit signatures of photoionized outflowing gas in their X-ray and UV spectra. Studying these outflows is important for a better understanding of the enrichment of the Inter Galactic Medium (IGM) as well as the physics of accretion of gas onto a super-massive black hole. 
\par
One of the best candidates to study these processes is NGC~5548. A recent XMM-{\it Newton} Reflection Grating Spectrometer (RGS) observation of NGC~5548 showed evidence for an X-ray warm absorber that has a nearly continuous distribution of ionization states (Steenbrugge et al. \cite{steen}). There is a pronounced increase in column density as a function of ionization parameter. The average outflow velocity measured in the X-ray band is consistent with the range of outflow velocities measured in the UV band, however, the spectral resolution in the X-ray band is not sufficient to resolve the individual velocity components observed in the UV band. The detection of the 1s-2p line of \ion{O}{vi} in X-rays allows for a direct comparison with the column density as determined from measurements of the 2s-2p lines by FUSE in the UV band (Brotherton et al. \cite{brotherton}). From this direct comparison, as well as the total column density derived from other lowly ionized ions, Arav et al. (\cite{arav03}) and Steenbrugge et al. (\cite{steen}) concluded that there is substantially more lowly ionized gas than has been claimed from previous UV observations. It was concluded that the X-ray and UV warm absorbers are different manifestations of the same phenomenon. Mathur, Elvis \& Wilkes (\cite{mathur}) already noted this possibility before the advent of high resolution X-ray spectroscopy. However, their model included only one ionization component and this is unrealistic for the observed NGC~5548 high resolution spectra.
\par
The difference in derived column densities in the UV band versus X-rays can be at least partially explained if the UV absorption lines do not cover the narrow emission line (NEL) region. The column densities derived from the UV lines are then only lower limits due to saturation effects, because these lines then locally absorb almost all the radiation from the continuum and the broad emission line (BEL) region (Arav et al. \cite{arav02}; Arav et al. \cite{arav03}). 
\par
The discrepancy in the \ion{O}{vi} column density mentioned above is based on non-simultaneous X-ray and UV observations and therefore it remains possible that the difference is due to variability in the absorber. The present {\it Chandra} observation was proposed to remedy this, by having simultaneous HST STIS, FUSE and {\it Chandra} observations. Unfortunately, due to technical problems FUSE was unable to observe at the scheduled time, so that no simultaneous \ion{O}{vi} observations were obtained. However, the spectral resolution of {\it Chandra} allows us to resolve the $-$1040 km~s$^{-1}$ component from the four other velocity components detected in the UV, for the strongest lines. This is a rather stringent test of the kinematic relation between both absorbers. 
\par
The long wavelength range of the LETGS allows us to map out the column density as a function of ionization parameter. Using only iron to avoid abundance effects we can sample about three orders of magnitude in ionization parameter from \ion{Fe}{vi} to \ion{Fe}{xxiv}. Due to the high signal-to-noise ratio, we are sensitive to changes in ionization parameter as small as 0.15 in log $\xi$. This is the first Seyfert 1 spectrum where the signal to noise is such that we are able to study in detail the absorption lines between 60 $-$ 100 \AA.
\par
The details of the observations and the data reduction are given in Sect. 2. In Sect. 3 we present the spectral data analysis. The short and long term spectral variability of the warm absorber are discussed in Sect. 4. In Sect. 5 we discuss our results. In Sect. 6 we discuss the ionization structure, the geometry and the mass loss through the outflow discussed in Sect. 7. The continuum time variability is discussed in detail in Kaastra et al. (\cite{kaastra04}). Limits on the spatial distribution of the X-ray emission are given by Kaastra et al. (\cite{kaastra03}).  The data were taken nearly simultaneously with HST STIS observations detailed by Crenshaw et al. (\cite{crenshaw}). 

\section{Observation and data reduction\label{sect:obs}}
NGC~5548 was observed for a full week with the {\it Chandra} High Energy Transmission Grating Spectrometer (HETGS) and the Low Energy Transmission Grating Spectrometer (LETGS) in January 2002. In Table~\ref{tab:obs} the instrumental setup and the exposure times are listed. 

\begin{table}
\begin{center}
\caption{The exposure time and the observation details for the {\it Chandra} and HST observations of NGC~5548. ACIS stands for the Advanced CCD Imaging Spectrometer and HRC for High Resolution Camera. The E140M grating covers the wavelength range between $1150 - 1730$ \AA, the E230M grating between $1607 - 3119$ \AA.}
\label{tab:obs}
\begin{tabular}{|l|l|l|l|}\hline
Instrument&detector&start date  &exposure (ks)\\\hline
HETGS     & ACIS-S &2002 Jan. 16& 154 \\
LETGS     & HRC-S  &2002 Jan. 18& 170 \\
LETGS     & HRC-S  &2002 Jan. 21& 170 \\\hline
HST STIS  & E140M  &2002 Jan. 22& 7.6 \\
HST STIS  & E230M  &2002 Jan. 22& 2.7 \\
HST STIS  & E140M  &2002 Jan. 23& 7.6 \\
HST STIS  & E230M  &2002 Jan. 23& 2.7 \\\hline
\end{tabular}
\end{center}
\end{table}

The LETGS data were reduced as described by Kaastra et al. (\cite{kaastra02}). The LETGS spans a wavelength range from $1-180$~\AA~with a resolution of 0.05 \AA~(full width half maximum, FWHM). The data were binned to 0.5 FWHM. However, due to Galactic absorption toward NGC~5548, the spectrum is heavily absorbed above 60 \AA, and insignificant above 100 \AA. In the present analysis we rebinned the data between $60 - 100$ \AA~by a factor of 2 to the FWHM of the instrument and ignore the longer wavelengths. 
\par
The HETGS data were reduced using the standard CIAO software version 2.2. The HETGS spectra consist of a High Energy Grating (HEG) spectrum which covers the wavelength range of $1.5-13$ \AA~with a FWHM of 0.012 \AA, and a Medium Energy Grating (MEG) spectrum which covers the wavelength range of $1.5-24$ \AA~with a FWHM of 0.023 \AA. The MEG and HEG data are binned to 0.5 FWHM. In the reduction of both HETGS datasets the standard ACIS contamination model was applied (see http://cxc.harvard.edu/caldb/about$_-$CALDB/index.html).
\par
The data shortward of 1.5 \AA~were ignored for all the instruments, as were the data above 11.5 \AA~for the HEG spectrum and above 24 \AA~for the MEG spectrum. The higher orders are not subtracted from the LETGS spectrum, but are included in the spectral fitting, i.e. the response matrix used includes the higher orders up to order $\pm$ 10, with the relative strengths of the different orders taken from ground calibration and in-flight calibrations of Capella. We did not analyze the higher order HETGS spectra, due to their low count rate. All the spectra were analyzed with the SPEX software (Kaastra et al. \cite{kaastrasp}), and the quoted errors are the 1 $\sigma$ rms uncertainties, thus $\Delta\chi^2$ = 2. 
\par
Kaastra et al. (\cite{kaastra02}) found a difference in the wavelength scale between the MEG and LETGS instrument of about 170~km~s$^{-1}$. This wavelength difference is within the absolute calibration uncertainty of the MEG instrument. The best line in NGC~5548 to test the wavelength calibration is the \ion{O}{vii} forbidden emission line, as it is found to be narrow. However, due to the noise in the MEG spectrum, the wavelength is not well determined, but it is consistent with the wavelength measured from the LETGS spectrum. We also looked for differences in the line centroid of the \ion{O}{viii} Ly$\alpha$ and \ion{O}{viii} Ly$\beta$ absorption line. However, as these lines are broadened, centroids are less accurate. These lines have consistent centroids for both instruments, and no systematic blue- or redshift is detected. 

\section{Spectral data analysis\label{sect:spec}}
In the online appendix C the spectrum between 1.5 \AA~$-$ 100 \AA~is shown. The X-ray spectrum is qualitatively similar to the spectra observed before by {\it Chandra} (Kaastra et al. \cite{kaastra00}; Kaastra et al. \cite{kaastra02}; Yaqoob et al. \cite{yaqoob}) and XMM-{\it Newton} (Steenbrugge et al. \cite{steen}). A large number of strong and weak absorption lines cover a smooth continuum. In addition, a few narrow and broadened emission lines are visible. Most absorption lines can be identified with lines already observed in these previous spectra or with predicted lines that, due to the increased sensitivity of the present spectra, become visible for the first time.
\par
The spectral model applied here is based on results obtained from earlier {\it Chandra} and XMM-{\it Newton} spectra of NGC~5548 (Kaastra et al. \cite{kaastra02}; Steenbrugge et al. \cite{steen}). The current spectrum is described by the same spectral components as in the earlier observations, albeit with different parameters. In the next subsections we discuss the results for the different components as obtained from our global fit to the spectra. These previous observations as well as the present observation have a continuum that is well fitted by a power-law and modified black body component {(Kaastra \& Barr \cite{kaastrab}) modified by Galactic extinction and cosmological redshift. The Galactic \ion{H}{i} column density was frozen to 1.65 $\times$ 10$^{24}$ m$^{-2}$ (Nandra et al. \cite{nandra}), as was the redshift of 0.01676 from the optical emission lines (Crenshaw \& Kraemer \cite{crenshaw99}). The continuum parameters for our best fit model, namely the one with three {\it slab} components (Sect.~\ref{sect:modela}, model A), are listed in Table~\ref{tab:cont}. For comparison, the continuum parameters for the earlier RGS observation are included (Steenbrugge et al. \cite{steen}).} 
\par
The high resolution spectra reveal a warm absorber (WA) modeled by combinations of {\it slab}, {\it xabs} or {\it warm} components (Sect.~\ref{sect:sep}), narrow emission lines (Sect.~\ref{sect:nel}~and~\ref{sect:feka}) and broadened \ion{O}{viii}, \ion{O}{vii} and \ion{C}{vi} emission lines (Sect.~\ref{sect:bel}).

\begin{figure}
 \resizebox{\hsize}{!}{\includegraphics[angle=-90]{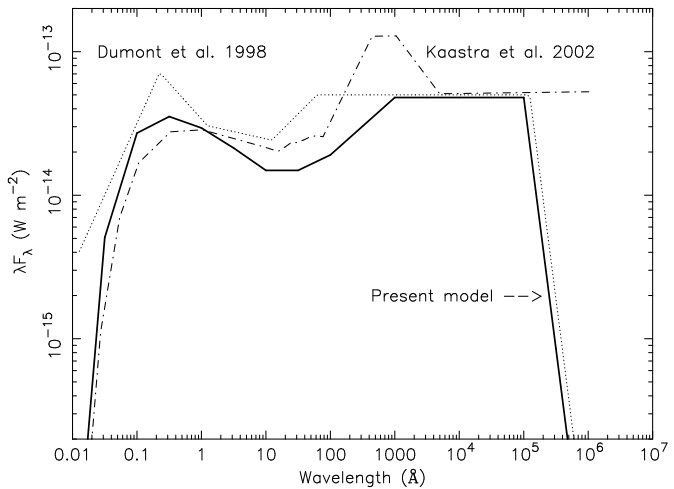}}
\caption{The SED used (solid line) in the present analysis and the analysis of the HST STIS data by Crenshaw et al. (\cite{crenshaw}). For comparison, the SED used by Kaastra et al. (\cite{kaastra02}) and Steenbrugge et al. (\cite{steen}) in earlier papers on NGC~5548 is plotted as the dot-dash line. The SED adopted by Dumont et al. (\cite{dumont}) is plotted as the dotted line.}
\label{fig:sed}
\end{figure}

\begin{table}
\caption{The best fit parameters for the continuum for the earlier RGS observation (Steenbrugge et al. \cite{steen}) and the present HETGS and LETGS observations of NGC~5548. The parameters for the HETGS and LETGS are for the best fit model including the three {\it slab} components with the outflow velocities frozen to the UV values.}
\label{tab:cont}
\begin{tabular}{lccc}\hline
Parameter & RGS       & HETGS             & LETGS              \\\hline
pl:norm$^{a}$  & 1.57 $\pm$ 0.03 & 0.90 $\pm$ 0.01 &  1.30 $\pm$ 0.01               \\
pl:lum$^{b}$ & 5.7 $\pm$ 0.1 & 3.62 $\pm$ 0.04 & 4.02 $\pm$ 0.03      \\
pl:$\Gamma$  & 1.77 $\pm$ 0.02 & 1.71 $\pm$ 0.01 & 1.88 $\pm$ 0.01 \\
mbb:norm$^{c}$  & 6.5 $\pm$ 0.7 & $<$ 400 & 3.5 $\pm$ 0.9               \\
mbb:$T$ (in eV) & 97 $\pm$ 6 & $-$  &  100 $\pm$ 10 \\
\end{tabular}\\
$^{a}$ At 1 keV in 10$^{52}$ ph s$^{-1}$ keV$^{-1}$. \\
$^{b}$ The luminosity as measured in the $2-10$ keV band in 10$^{36}$ W. \\
$^{c}$ norm = emitting area times square root of electron density in 10$^{32}$ m$^{1/2}$.
\end{table}

\subsection{Warm absorber models\label{sect:sep}}

In our modeling of the warm absorber we use combinations of three different spectral models for the absorber. The most simple is the {\it slab} model (Kaastra et al. \cite{kaastra02}), which calculates the transmission of a slab of a given column density and a set of ion concentrations. Both the continuum and the line absorption are taken into account. As in the other absorption models, the lines are modeled using Voigt profiles. Apart from the ionic column densities, the average outflow velocity $v$ and the Gaussian r.m.s. velocity broadening $\sigma$ are free parameters of the model. The last two parameters have the same value for each ion. The slab is assumed to be thin in the sense that $e^{-\tau}$ gives a fair description of its transmission. 
\par
The second model ({\it xabs}) is similar to the {\it slab} model, however now the ionic column densities are coupled through a set of photoionization balance calculations (Kaastra et al. \cite{kaastra02}). As a result, instead of the individual ionic concentrations we now have the ionization parameter $\xi$ = $L$/$n_er^2$ and the elemental abundances as free parameters. The set of photoionization balance calculations was done using the XSTAR code (Kallman \& Krolik \cite{kallman}), with the SED as shown in Fig.~\ref{fig:sed}. The same SED was used by Crenshaw et al. (\cite{crenshaw}) in the analysis of the simultaneous HST STIS data. As there is no evidence of a blue bump this was not included in the currently used SED. This is the main difference with the SED used by Kaastra et al. (\cite{kaastra02}) and Steenbrugge et al. (\cite{steen}), which is plotted as the dot-dash line. For the X-ray part of the spectrum we used the continuum model as measured from the present LETGS spectrum, instead of the continuum obtained from the 1999 LETGS spectrum. For the high energy part of the SED we used a power law component as measured in our {\it Chandra} spectrum but with an exponential cut-off at 130 keV, consistent with the Beppo{\it SAX} data (Nicastro et al. \cite{nicastro}). As reflection components are time variable and the LETGS has insufficient high energy sensitivity to model it properly, we did not include any reflection component in our SED. Between 50 and 1000 \AA~we used a power-law and at even longer wavelength used the energy index obtained by Dumont et al. (\cite{dumont}) of 2.5.
\par
The third method used is the {\it warm} model, which is a model for a continuous distribution of the d$N_{\rm H}$/d$\xi$ as a function of $\xi$. At each value of $\xi$, the transmission is calculated using the {\it xabs} model described above. In our implementation of this model $\xi {\rm d}N_{\rm H}/{\rm d}\xi$ is defined at a few (in our case two) values of $\xi$. At the intermediate points the logarithm of this function is determined by cubic spline interpolation on the log $\xi$ grid. This determines $N_{\rm H}$ as a function of $\xi$. Integration over $dN_{\rm H}$/d$\xi$ between the lowest and the highest values of $\xi$ ($\xi_1$ respectively $\xi_2$) yields the total ionic column densities. In our model, as we have taken only two points, the total hydrogen column density is therefore a power-law function of ionization. The {\it warm} model correctly incorporates the fact that ions are formed at a range of ionization parameters. Free parameters of this model are $\xi_1$, $\xi_2$, $\xi_1$~d$N_{\rm H}~(\xi_1)$/d$\xi$, $\xi_2$~d$N_{\rm H}~(\xi_2)$/d$\xi$, and similar to the {\it xabs} model the outflow velocity $v$, velocity broadening $\sigma$ and the elemental abundances. 
\par
Finally, in all our models ({\it slab}, {\it xabs} and {\it warm}) we changed the wavelength for the \ion{O}{v} 1s${^2}$2s${^2}$ - 1s2s${^2}$2p $^{1}$P$_1$ X-ray line from 22.33 \AA~(the HULLAC value) to 22.374~\AA~(Schmidt et al. \cite{schmidt}) as measured at the University of California EBIT-I electron beam ion trap. This recent value was not yet available for our earlier analysis of the RGS data (Steenbrugge et al. \cite{steen}).

\subsection{Spectral fits with warm absorber models}

\subsubsection{Warm absorber model A: column densities with UV velocity structure\label{sect:modela}}
Our first method (dubbed model A here) allows us to further ascertain that the X-ray warm absorber and the UV absorber are the same phenomenon, a major goal for proposing this observation. We fitted the warm absorber in the X-ray spectra with the outflow velocities frozen to those measured in the UV spectra. 
\par
\begin{figure}
 \resizebox{\hsize}{!}{\includegraphics[angle=-90]{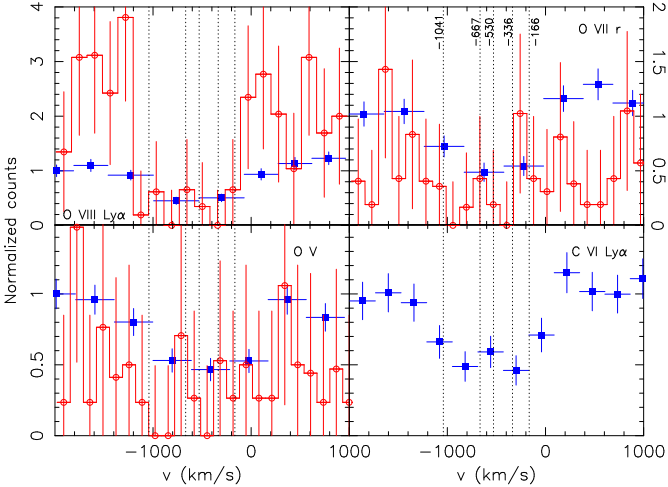}}
\caption{The MEG (open circles) and LETGS (filled squares) line profiles for the deepest line of \ion{O}{viii}, \ion{O}{vii}, \ion{O}{v} and \ion{C}{vi}. The five outflow velocities measured in the UV are indicated by the dotted line. No correction was made for possible blending.}
\label{fig:profile}
\end{figure}

We implemented this by modeling the WA using three {\it slab} components, with the outflow velocities frozen. These outflow velocities were frozen to $-1040$ km~s$^{-1}$, $-530$ km~s$^{-1}$ and $-160$ km~s$^{-1}$, corresponding to UV velocity components 1, the average of $2-4$, and 5, respectively. Note that the velocities of components $2-4$ ($-$667, $-$530 and $-$336~km~s$^{-1}$, respectively) are too close to be separated in our X-ray spectra. In the fit we left the velocity broadening of each of the three components a free parameter. Allowing for three velocity components instead of one, substantially improved the fit from $\chi^2$ = 2916 to 2441 for 2160 degrees of freedom (only LETGS is quoted here). However, the absorption lines are still only partially resolved, leading to strongly correlated errors for the derived column densities of the velocity components. Only for the six ions with the deepest lines (\ion{C}{vi}, \ion{O}{v}, \ion{O}{vii}, \ion{O}{viii}, \ion{Ne}{ix} and \ion{Si}{xiii}) we measure column densities with meaningful error ranges (Table~\ref{tab:3v}). For these six ions the velocity structure of the X-ray warm absorber closely resembles the UV velocity structure. Fig.~\ref{fig:profile} shows the line profiles as a function of outflow velocity for four lines of the ions listed in Table~\ref{tab:3v}. The HEG spectrum shows most clearly substructure in the \ion{Mg}{xii} Ly$\alpha$ absorption lines (Fig.~\ref{fig:shift1} in Appendix C online).

\begin{table}
\begin{center}
\caption{Parameters for six ions measured using model A. The first row gives the component number as listed in the UV. The second row lists the outflow velocity $v$, which was frozen to the UV values, the third row lists the velocity broadening $\sigma_{\rm X-ray}$ as derived from the X-ray spectra. In the fourth row the velocity broadening $\sigma_{\rm UV}$ = FWHM/2.35, where FWHM is the measured and resolved Full Width Half Maximum of the UV absorption lines (Crenshaw et al. \cite{crenshaw}). All are in km~s$^{-1}$. For the $-530$~km~s$^{-1}$ component we added the velocity broadening on component 3 and 4. We list the logarithms of the column densities in m$^{-2}$. For comparison we also list the \ion{C}{iv} and \ion{N}{v} column densities as measured by Crenshaw et al. (\cite{crenshaw}).}
\label{tab:3v}
\begin{tabular}{lcccc}\hline
comp.        & 1              & $2 - 4$        & 5              &            \\
$v$  &  $-1040$       & $-530$         & $-160$         & \\
$\sigma_{\rm X-ray}$  &  40 $\pm$ 5    & 100 $\pm$ 15   & 90 $\pm$ 13 &  \\
$\sigma_{\rm UV}$  &  94 $\pm$ 8  & 140 $\pm$ 15   & 26 $\pm$ 6     &  \\
ion          &                &                &                & total \\\hline
\ion{C}{iv}  & 18.05 $\pm$ 0.05 & 18.66 $\pm$ 0.02 & 17.76 $\pm$ 0.03 & 18.8 \\
\ion{C}{vi}  & 20.2 $\pm$ 0.6 & 21.4 $\pm$ 0.2 & 20.8 $\pm$ 0.4 & 21.5 \\
\ion{N}{v}   & 18.44 $\pm$ 0.02 & 19.24 $\pm$ 0.02 & 18.16 $\pm$ 0.03  & 19.3 \\
\ion{O}{v}   & 20.0 $\pm$ 0.5 & 20.5 $\pm$ 0.3 & 20.2 $\pm$ 0.6 & 20.8 \\
\ion{O}{vii} & 21.3 $\pm$ 0.5 & 21.4 $\pm$ 0.3 & 20.3 $\pm$ 0.7 & 21.7 \\
\ion{O}{viii}& 22.2 $\pm$ 0.1 & 21.5 $\pm$ 0.3 & 21.9 $\pm$ 0.3 & 22.4 \\ 
\ion{Ne}{ix} & 20.5 $\pm$ 0.9 & 20.4 $\pm$ 0.6 & 20.7 $\pm$ 0.8 & 21.0 \\
\ion{Si}{xiii}&20.8 $\pm$ 0.6 & 20.6 $\pm$ 0.6 & 20.5 $\pm$ 1.1 & 21.1 \\
\end{tabular}
\end{center}
\end{table}

\subsubsection{Warm absorber model B: column densities with simple velocity structure\label{sect:modelb}}
\noindent{\bf Free velocity broadening}

\begin{table*}
\begin{center}
\caption{The best fit column densities using model B for the ions for which the outflow velocity and velocity broadening are well constrained. The outflow velocity $v$ and the r.m.s. velocity broadening $\sigma$ are listed as well as the ionization parameters $\xi$ and $U$ for which the ion has its maximum column density. The column densities are from the fit including the broad emission lines (Sect.~\ref{sect:bel}).}
\label{tab:1v}
\begin{tabular}{lclcrr}\hline
ion            & log $N_{\rm ion}$ &  $- v$          & $\sigma$    & log $\xi$     & log $U$  \\
               & (m$^{-2}$)      & (km s$^{-1}$) & (km s$^{-1}$)& (10$^{-9}$ W m) &          \\\hline
\ion{C}{v}     & 21.2 $\pm$ 0.5  & 530 $\pm$ 90  & 185 $\pm$ 70 & 0.10             & $-$1.50  \\
\ion{C}{vi}    & 21.6 $\pm$ 0.1  & 480 $\pm$ 60  & 180 $\pm$ 35 & 1.00             & $-$0.60  \\
\ion{N}{vi}    & 21.2 $\pm$ 0.2  & 550 $\pm$ 130 & 200 $\pm$ 120& 0.50             & $-$1.10  \\
\ion{N}{vii}   & 21.5 $\pm$ 0.2  & 320 $\pm$ 110 & 390 $\pm$ 120& 1.35            & $-$0.20  \\
\ion{O}{v}     & 20.8 $\pm$ 0.2  & 530 $\pm$ 120 & 280 $\pm$ 100& $-$0.20          & $-$1.80  \\
\ion{O}{vi}    & 20.6 $\pm$ 0.2  & 380 $\pm$ 150 & 160 $\pm$ 110& 0.30             & $-$1.30  \\
\ion{O}{vii}   & 22.18 $\pm$ 0.03& 590 $\pm$ 60  & 150 $\pm$ 20 & 0.95            & $-$0.65  \\
\ion{O}{viii}  & 22.53 $\pm$ 0.04& 540 $\pm$ 60  & 130 $\pm$ 20 & 1.65            & 0.05    \\
\ion{Ne}{ix}   & 21.2 $\pm$ 0.3  & 660 $\pm$ 170 & 210 $\pm$ 110& 1.50             & $-$0.10  \\
\ion{Ne}{x}    & 21.9 $\pm$ 0.2  & 830 $\pm$ 140 & 290 $\pm$ 60 & 2.05            & 0.45    \\
\ion{Mg}{ix}   & 21.0 $\pm$ 0.3  & 560 $\pm$ 70  & 55 $\pm$ 30  & 1.00             & $-$0.60  \\
\ion{Mg}{xii}  & 21.3 $\pm$ 0.1  & 680 $\pm$ 120 & 430 $\pm$ 120& 2.35            & 0.75    \\
\ion{Si}{viii} & 20.2 $\pm$ 0.2  & 620 $\pm$ 210 & 330 $\pm$ 190& 0.40             & $-$1.20  \\
\ion{Si}{x}    & 20.2 $\pm$ 0.2  & 790 $\pm$ 150 & 340 $\pm$ 150& 1.05            & $-$0.55  \\
\ion{Si}{xi}   & 20.0 $\pm$ 0.2  & 810 $\pm$ 180 & 350 $\pm$ 150& 1.50             & $-$0.10  \\
\ion{Si}{xiii} & 21.0 $\pm$ 0.1  & 660 $\pm$ 280 & 80 $\pm$ 40  & 2.20             & 0.60    \\
\ion{Si}{xiv}  & 21.0 $\pm$ 0.3  & 880 $\pm$ 120 & 60 $\pm$ 30  & 2.60             & 1.00    \\
\ion{S}{xi}    & 20.0 $\pm$ 0.3  & 560 $\pm$ 200 & 310 $\pm$ 170& 1.20             & $-$0.30  \\
\ion{S}{xii}   & 20.2 $\pm$ 0.2  & 620 $\pm$ 300 & 450 $\pm$ 200& 1.45            & $-$0.15  \\
\ion{Fe}{xvii} & 20.3 $\pm$ 0.4  & 740 $\pm$ 110 & 50 $\pm$ 30  & 1.80            & +0.20    \\
\end{tabular}
\end{center}
\end{table*}

Using model A we can only derive accurate column densities for six ions. In order to obtain column densities for other ions we need to reduce the number of free parameters. To do so we use a model with one outflow velocity component, leaving the velocity broadening as well as the outflow velocity free. This allows us to obtain column densities, outflow velocities and velocity broadening for the ions with stronger lines. We first made a fit with a single {\it slab} component. This {\it slab} component has a single outflow velocity and velocity broadening for all the ions. After this fit is obtained, we freeze all the parameters in this {\it slab} component. 
\par
\begin{figure}
 \resizebox{\hsize}{!}{\includegraphics[angle=-90]{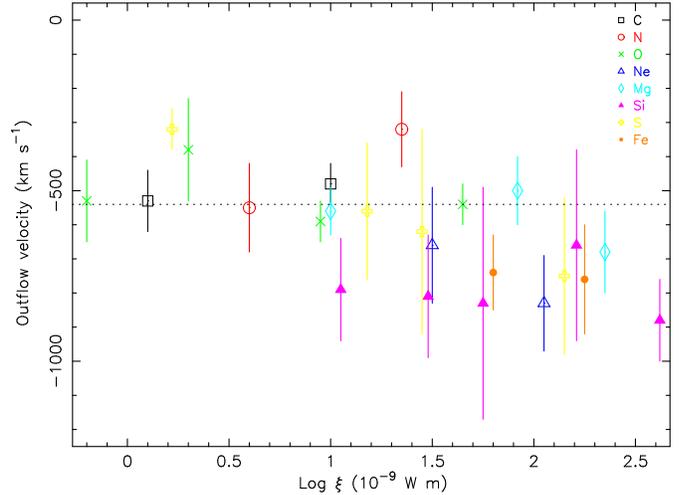}}
\caption{The measured outflow velocity versus the logarithm of the ionization parameter. The dotted line indicates the $-$530~km~s$^{-1}$ outflow velocity, the dominant component in the UV absorber. A trend toward higher outflow velocities for higher ionized ionization parameters is observed.}
\label{fig:belzv}
\end{figure}

A second {\it slab} component is then added containing only a single ion, leaving the column density, outflow velocity and velocity broadening as free parameters. The column density for this particular ion is frozen to zero in the first {\it slab} component, and the spectra are fit again in order to fine tune the column density, outflow velocity and velocity broadening for each ion. Each ion was cycled through the second {\it slab} component in this manner, and the results for those ions which have well determined outflow velocity and velocity broadening are listed in Table~\ref{tab:1v}. In this Table we also list the ionization parameter for which the column density of the particular ion peaks. We list $\xi$ = $L$/$n_er^2$ expressed in 10$^{-9}$ W m, where $L$ is the energy luminosity, $n_e$ is the electron density, $r$ the distance. We also list $U$ = $Q_{\rm tot}$/(4$\pi$$r^2$$c$$n_e$), with $Q_{\rm tot}$/(4$\pi$$r^2$) the total photon flux for photons with an energy greater than 13.6 eV. This Table indicates a weak correlation between outflow velocity and the ionization of the ion, which is shown in Fig.~\ref{fig:belzv}. As a side effect, leaving the outflow velocity free potentially allows to correct for any remaining wavelength inaccuracies.

\noindent{\bf Fixed velocity broadening}
\par
As it is not possible to determine the outflow velocity and broadening accurately for all ions, we also fitted the spectra using a fixed velocity broadening to further reduce the number of free parameters. The velocity broadening was fixed to 140~km~s$^{-1}$ (average $\sigma$ for \ion{O}{vii} and \ion{O}{viii}) and 70~km~s$^{-1}$ (average $\sigma$ for \ion{Si}{xiii} and \ion{Si}{xiv}). Those two values were chosen to study the influence of saturation and velocity broadening on the determined column density. In this fit the outflow velocity was fixed to either the derived value (see Table~\ref{tab:1v}) or a value between $-$400 and $-$800~km~s$^{-1}$, depending on the ionization parameter of the ion. This method allows us to obtain an accurate column density for the less abundant ions. The results of this fit are listed in the appendix A as Table~\ref{tab:dnh}. As in Table~\ref{tab:1v} we list the ionization parameter for which the column density of the particular ion peaks. For several ions we find a higher column density and larger velocity broadening in Table~\ref{tab:1v}, which can be explained by the fact that a velocity broadening of 140 km s$^{-1}$ does not include the full blend due to the five different velocity components. 

\subsubsection{Warm absorber model C: multiple separate ionization components\label{sect:modelc}}

In our previous models A and B we determined ionic column densities independently from the photoionization balance. A disadvantage of this method is that the column densities of the many ions with small abundances are poorly constrained, while the combined effect of these ions may still be noticeable in the spectrum. Therefore we have tried spectral fits with three {\it xabs} components.
\par
The fit with three {\it xabs} components is not statistically acceptable. As in previous papers the iron abundance is measured to be overabundant for lower ionization ions (Kaastra et al. \cite{kaastra02}; Blustin et al. \cite{blustin}; Steenbrugge et al. \cite{steen}; Netzer et al. \cite{netzer}). A possible explanation was given by Netzer et al. (\cite{netzer}). Namely, the ionization balance for M-shell iron in low temperature, photoionized plasmas depends strongly on the dielectronic recombination rates. These rates are sometimes poorly known due to resonance effects at low energies, and are thus not correctly incorporated in the photoionization balance codes that are used to predict the ionic column densities. In recent articles Kraemer, Ferland \& Gabel (\cite{kraemer}) and Netzer (\cite{netzer1}) estimate that the ionization parameter for the iron ions that produce the UTA can change by as much as a factor of two or more. An increase of the effective ionization parameter $\xi$ by a factor $\sim$ 2 for \ion{Fe}{vii} - \ion{Fe}{xiii} indeed would bring these ions more in line with the trend observed for the other elements. 
\par
Fitting iron simultaneously with the other elements will bias the ionization parameter measured. A solution is to decouple iron from the other elements. Since however we need at least three {\it xabs} components, this would imply doubling of the number of free parameters to six {\it xabs} models. Another option is to fit instead iron separately with a {\it slab} model. The results of this fit are listed in Table~\ref{tab:xabs}. However, as iron has the largest ionization range and is the only element that samples both the highest ionized and lowest ionized absorber, both options severely limit the ionization range we are able to detect unbiased. The error bars derived for the fit without iron are much larger; and the ionization parameters measured should not be compared with those listed in our earlier RGS analysis, as we included iron in the fit. Forcing a large velocity broadening to fit all the outflow velocity components, and fitting the iron ions separately with a {\it slab} model, we find a decent fit for three ionization components ($\chi^2$ = 3435 for 2758 degrees of freedom). Adding more {\it xabs} components leads to highly correlated errors and fitting results. We list the results of this fit in Table~\ref{tab:xabs}.

\begin{table}
\begin{center}
\caption{The best fit results for a model with three {\it xabs} components fitting all elements but iron, which was fit separately with a {\it slab} model (model C). To fit the full blend we froze the velocity broadening to 200 km s$^{-1}$ and the outflow velocity to $-$ 530 km s$^{-1}$.}
\label{tab:xabs}
\begin{tabular}{llll}\hline
Comp.             & A             & B           & C     \\
LETGS:            &               &    &    \\\hline
log N$_{\rm H}$ (m$^{-2}$) & $<$ 25.9 & $<$ 25.8 & 24.56$\pm$0.09 \\
log $\xi$ (10$^{-9}$ W m)  & $<$ 5    & 2.1$\pm$0.6 & 0.8$\pm$0.1\\
LETGS, HEG and MEG:&                  &             &     \\\hline
log N$_{\rm H}$ (m$^{-2}$) & $<$ 28   & 25.43$\pm$0.06&24.55$\pm$0.08 \\
log $\xi$ (10$^{-9}$ W m)  & $<$ 5    & 2.17$\pm$0.05& 0.80$\pm$0.09\\
\end{tabular}
\end{center}
\end{table}

Considering the error bars, due to the exclusion of iron in the fit, the model with different ionization parameters is only constrained for two ionization components (B and C) using the full observation. In order to further reduce the number of free parameters, and thus to obtain a better constrained fit as well as to test whether the ionization parameter distribution is continuous, we next test a model with a continuous ionization structure.

\subsubsection{Warm absorber model D: continuous ionization structure\label{sect:modeld}}

\begin{figure}
 \resizebox{\hsize}{!}{\includegraphics[angle=-90]{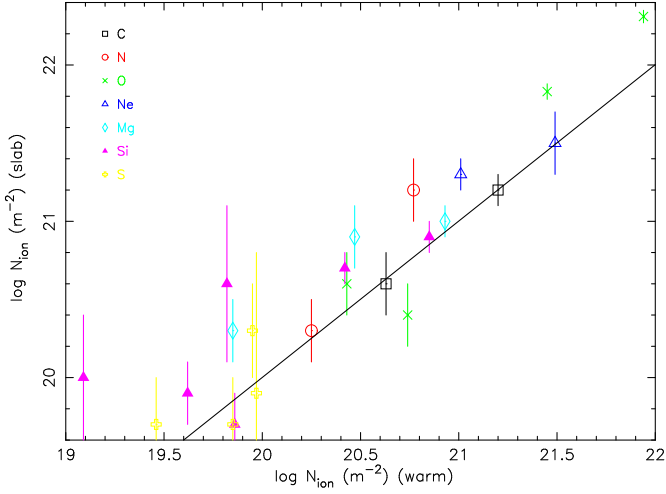}}
\caption{Comparison between the ionic column densities calculated with the {\it warm} model (x-axis is model D) and the ionic column densities obtained with the {\it slab} model (y-axis is model B). The solid line represents the solution where the predicted {\it warm} and measured {\it slab} column densities are equal.}
\label{fig:comp}
\end{figure}

Steenbrugge et al. (\cite{steen}) found a power law like distribution of deduced hydrogen column density versus ionization parameter. In order to test whether this apparent power law distribution can be produced indeed by an underlying power law, we have fitted the spectrum directly to such a distribution using the {\it warm} model (Sect.~\ref{sect:sep}). Similar to model C, we get a poor fit if iron is included. Modeling all iron ions using a {\it slab} model and all remaining elements with the {\it warm} model, we obtain a good fit to the data, namely $\chi^2$ = 3305 for 2754 degrees of freedom, assuming solar abundances. Table~\ref{tab:warm} lists the best fit parameters using the {\it warm} model. In this fit the best fit outflow velocity is $-$530~km~s$^{-1}$ and the measured velocity broadening is 140~km~s$^{-1}$. The derived column density distribution is in excellent agreement with the results obtained from model B (see Fig.~\ref{fig:comp}).

\begin{table}
\begin{center}
\caption{The best fit values for model D, using one {\it warm} component and fitting iron separately as a {\it slab} component.}
\label{tab:warm}
\begin{tabular}{cc}\hline
log $\xi$ (10$^{-9}$ W m) & $\xi$d$N_{\rm H}$/d$\xi$  \\\hline
$-$0.1                    & (6.3 $\pm$ 1.1) $\times$ 10$^{23}$ m$^{-2}$    \\
3.5                       & (1.6 $\pm$ 0.3) $\times$ 10$^{26}$ m$^{-2}$ 
\end{tabular}
\end{center}
\end{table}

In Fig.~\ref{fig:comp} we compare the ionic column densities predicted by model D with the column densities measured by model B. The correlation between the measured and predicted column density is rather tight, with a measured slope of 1.10 $\pm$ 0.07. However, one also notes that the measured column densities using model B are higher than for model D. This is due to the fact that with method B we optimized the outflow velocity. This is not possible with model D, where the outflow velocity and velocity broadening are tied for all ions. 

\subsection{Reliability of column density estimates\label{sect:id}}

In Table~\ref{tab:dnh}, in appendix A we list all the ions that we fitted with the {\it slab} model (model B). As the {\it slab} model fits both for the absorption lines and the corresponding edges, some column density determinations could be dominated by an edge measurement, rather than by absorption lines. Measurements dominated by an edge, although resulting in quite accurate column densities are highly dependent on the correct calibration of the instrument and the continuum fit, and thus potentially have a large systematic error. A further uncertainty is introduced by unknown blending at the longer wavelengths (i.e. above 60 \AA), as the atomic data for lines produced at this wavelength region are sometimes poorly known. In Appendix B we discuss those ions for which the column density maybe affected by the above effects.

\subsection{Narrow emission lines\label{sect:nel}}

\begin{table}
\caption{Narrow emission lines. The Equivalent Width (EW) as measured  with the LETGS, except for the \ion{Mg}{xi}, \ion{Al}{xiii} and \ion{Si}{xiii} forbidden lines (MEG), and flux are listed. In the last column the ionization parameter, $\xi$, where the ion has the highest column density is given. Forbidden lines are indicated by f, intercombination lines by i. Below the line the \ion{C}{v} and \ion{O}{vii} RRC are listed. The rest wavelengths are taken from Drake (\cite{drake}).}
\label{tab:nel}
\begin{tabular}{lcccc}\hline
line         & wavelength & EW       & flux      &  log $\xi$$^{a}$ \\ 
             & (\AA)      & (m\AA)   & (ph~m$^{-2}$~s$^{-1}$) & \\\hline
\ion{C}{v}f  & 41.435     & 283 $\pm$ 170 & 1.0 $\pm$ 0.6   & 0.1   \\
\ion{N}{vi}f & 29.518     & 37 $\pm$ 19   & 0.26 $\pm$ 0.13 & 0.6 \\
\ion{N}{vi}i & 29.082     & $<$ 17        & $<$ 0.1         & 0.6 \\
\ion{O}{vii}f& 22.093     & 103 $\pm$ 14  & 0.88 $\pm$ 0.12 & 0.95\\
\ion{O}{vii}i& 21.804     & $<$ 5         & $<$ 0.04        & 0.95\\
\ion{Ne}{ix}f & 13.690    & 28 $\pm$ 10   & 0.14 $\pm$ 0.05 & 1.5 \\
\ion{Ne}{ix}i & 13.566    & 12 $\pm$ 9    & 0.05 $\pm$ 0.03 & 1.5 \\
\ion{Mg}{xi}f&  9.314     & $<$ 4         & $<$ 0.02        & 1.9 \\
\ion{Al}{xii}f& 7.864     & 4 $\pm$ 2     & 0.04 $\pm$ 0.02 & 2.05\\
\ion{Si}{xiii}f& 6.739    & 5 $\pm$ 2.7   & 0.03 $\pm$ 0.02 & 2.2 \\\hline
\ion{C}{v}   & 31.63   & 23 $\pm$ 16   & 0.08 $\pm$ 0.06 & 0.1 \\   
\ion{O}{vii} & 16.77   & 51 $\pm$ 32   & 0.08 $\pm$ 0.05 & 0.95\\
\end{tabular}\\
$^{a}$ Given in 10$^{-9}$ W m.
\end{table} 

Our spectrum shows the presence of a few narrow emission lines. Table~\ref{tab:nel} lists the measured strength or upper limits of forbidden and intercombination lines of several ions as well as the Radiative Recombination Continua (RRC) observed in the spectra. For all narrow emission lines, with the exception of the \ion{O}{vii} and \ion{Ne}{ix} forbidden lines, the wavelength was frozen to the value in the restframe of NGC~5548. For the \ion{O}{vii} and \ion{Ne}{ix} forbidden lines we determined blueshifts of $-$0.009 $\pm$ 0.004 \AA~and $-$0.005 $\pm$ 0.007 \AA, corresponding to an outflow velocity of $-$150 km~s$^{-1}$ and $-$175~km~s$^{-1}$, respectively. These blueshifts, however, become negligible if instead, the redshift of 0.01717 based upon the 21 cm line (Crenshaw \& Kraemer \cite{crenshaw99}) is assumed.
\par
The intercombination lines as well as the RRC's are undetectable or very weak. In particular the \ion{O}{vii} intercombination line is not detected, likely due to blending by the absorption line of \ion{O}{vi} at 21.79 \AA~rest wavelength. The forbidden \ion{O}{vii} line seems double-peaked in the MEG spectrum (see online appendix C Fig.~\ref{fig:shift1}), and is poorly fit by a single Gaussian line. However, the LETGS spectrum and the earlier MEG spectrum of 2000 do not show this line profile. None of the other forbidden lines show any broadening or a double-peak profile. We thus conclude that the MEG \ion{O}{vii} forbidden line shape is due to noise. 

\subsection{Fe K$\alpha$\label{sect:feka}}
The narrow Fe K$\alpha$ emission line is clearly seen in the HEG spectrum (see online appendix C Fig.~\ref{fig:shift1}). The flux of the line is (0.24 $\pm$ 0.08) ph~m$^{-2}$~s$^{-1}$. Our results are in good agreement with the EPIC results presented by Pounds et al. (\cite{pounds}) and the earlier HEG data presented by Yaqoob et al. (\cite{yaqoob}) (Table~\ref{tab:fek}). The Fe K$\alpha$ emission line is also discussed by Yaqoob \& Padmanabhan (\cite{yaqoob2}). There is no evidence for a broadened Fe K$\alpha$ emission line, consistent with the results obtained by Yaqoob et al. (\cite{yaqoob}) and Pounds et al. (\cite{pounds}).

\begin{table}
\begin{center}
\caption{Parameters of the Fe K$\alpha$ line from the present HEG spectrum (2002), previous HEG spectrum (2000, Yaqoob et al. \cite{yaqoob}), and EPIC spectrum (2001, Pounds et al. \cite{pounds}). The FWHM is given in km~s$^{-1}$, the flux in ph m$^{-2}$ s$^{-1}$.}
\label{tab:fek}
\begin{tabular}{lccc}\hline
           & HEG (2000)       & EPIC (2001)    & HEG (2002) \\
E (keV)    & 6.402 $\pm$ 0.026& 6.39 $\pm$ 0.02& 6.391 $\pm$ 0.014\\
EW (eV)    & 133 $\pm$ 50     & 60 $\pm$ 15    & 47 $\pm$ 15 \\
FWHM       & 4515 $\pm$ 2650& 6500 $\pm$ 2200 & 1700 $\pm$ 1500 \\
flux       & 0.36 $\pm$ 0.16& 0.38 $\pm$ 0.10 & 0.24 $\pm$ 0.08\\
\end{tabular}
\end{center}
\end{table}

\subsection{Broad emission lines\label{sect:bel}}
\begin{figure}
 \resizebox{\hsize}{!}{\includegraphics[angle=-90]{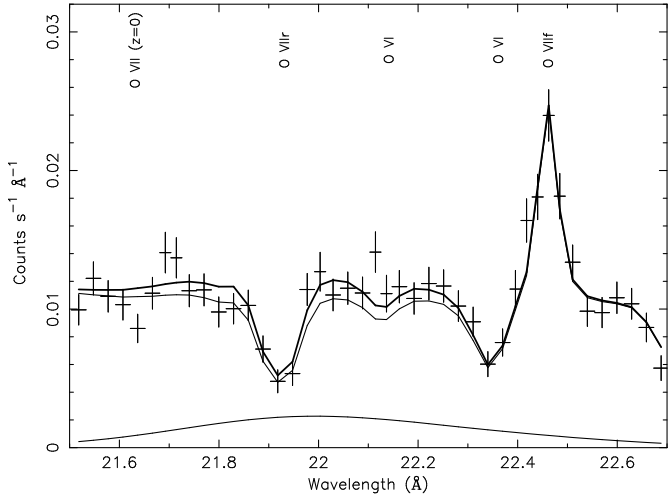}}
\caption{Detail of the LETGS spectrum showing the fit with (thick line) and without (thin line) a broad emission line for the \ion{O}{vii} resonance line. The profile of the broadened emission line is also plotted.}
\label{fig:broad}
\end{figure}

Kaastra et al. (\cite{kaastra02}) detected a 3$\sigma$ significant relativistically broadened emission line for \ion{N}{vii} Ly$\alpha$ and \ion{O}{viii} Ly$\alpha$, and a non-relativistically broadened emission line for \ion{C}{vi} Ly$\alpha$. Ogle et al. (\cite{ogle}) detect a broadened \ion{C}{vi} Ly$\alpha$ line with a FWHM of 1100 km s$^{-1}$ in NGC 4051. No relativistically or non-relativistically broadened emission lines were detected in the high flux state RGS spectrum (Steenbrugge et al. \cite{steen}). The LETGS and MEG spectra show clear, broad excess emission centered on some of the deeper absorption lines, although none show the asymmetric line profile of a relativistically broadened line (Fig. \ref{fig:broad}). We fit these broad excesses with Gaussians. Due to the noise in the data, the broadness of these features, and the rather low contrast with the continuum, we fixed the width of these Gaussians to 8000 km~s$^{-1}$ (Arav et al. \cite{arav02}) as observed for the broadest component of the \ion{C}{iv} and Ly$\alpha$ broad emission lines in the UV. The wavelength was fixed to the rest wavelength, i.e. assuming no outflow velocity. Table~\ref{tab:bel} lists the broad emission line fluxes for the LETGS and MEG spectra. A detail of the LETGS spectrum of the \ion{O}{vii} triplet broadened line is shown in Fig.~\ref{fig:broad}.

To ascertain the significance of these broad emission lines, we checked  the lines at shorter wavelengths separately in the LETGS and MEG spectra. The line fluxes are consistent within the error bars, between the MEG and LETGS observations. Inclusion of these broad emission lines in the model alters the derived column densities for the different ions in the warm absorber. Therefore the differences between the model with and without a broad emission line is not equal to the flux of this line (see Fig.~\ref{fig:broad}). As a result the column densities quoted in this paper, which are based on the fit with the broad emission lines included, should not be compared directly with column densities derived from earlier observations as small continuum mismatches are likely to remain in those spectra (e.g. Kaastra et al. \cite{kaastra02}; Steenbrugge et al. \cite{steen}) in a search for variability of the warm absorber. As an example, the logarithm of the derived column density for \ion{O}{vii} (using one {\it slab} component) in a model without a broad emission line is 21.72 m$^{-2}$, compared to 22.18 m$^{-2}$ if the broad emission lines are included. The broad emission lines are more easily detected than in previous observations, this partly results from the low continuum flux level, and thus the larger contrast during the present observation. Further discussion on these broad emission lines will be given in a forthcoming paper (Steenbrugge et al. \cite{steen04}).

\begin{table}
\caption{Flux of the broad emission lines, from the simultaneous LETGS, MEG and HEG fit. The wavelength was frozen to the rest wavelength of the line, while the FWHM was frozen to 8000 km~s$^{-1}$. The lines for which upper limits are detected were not used in the further analysis of the data.}
\label{tab:bel}
\begin{tabular}{llll}\hline
ion                           & $\lambda$  & flux                  & EW   \\
                              & (\AA)      & (ph~m$^{-2}$~s$^{-1}$) & (m\AA) \\\hline
\ion{C}{v} f$^{a}$ & 41.421  & 1.0 $\pm$ 0.6   & 300 $\pm$ 180 \\
\ion{C}{v} 1s$^2$-1s2p $^1$P$_1$          & 40.268  & $<$ 0.6         & $<$ 230      \\
\ion{C}{vi} 1s-2p (Ly$\alpha$)            & 33.736  & 0.5 $\pm$ 0.2   & 150 $\pm$ 80  \\
\ion{N}{vii} 1s-2p (Ly$\alpha$)           & 24.781  & $<$ 0.05        & $<$ 70       \\
\ion{O}{vii} triplet$^{b}$                & 21.602  & 0.56 $\pm$ 0.13  & 130 $\pm$ 40  \\
\ion{O}{viii} 1s-2p (Ly$\alpha$)          & 18.969  & 0.4 $\pm$ 0.2   & 60 $\pm$ 30   \\
\ion{O}{vii} 1s$^2$-1s3p $^1$P$_1$        & 18.627  & 0.19 $\pm$ 0.07  & 70 $\pm$ 30   \\
\end{tabular}\\
$^{a}$ This coincides with the instrumental C-edge, therefore its detection is less certain due to possible calibration uncertainties.\\
$^{b}$ The resonance intercombination and forbidden lines significantly overlap, so separate measurements are difficult. \\
\end{table}

\section{Timing analysis}
Kaastra et al. (\cite{kaastra04}) give a detailed analysis of the lightcurve in different energy bands. Here we focus on the possible differences in the warm absorber due to flux variations.

\subsection{Short term variability\label{sect:var}}

\begin{figure}
 \resizebox{\hsize}{!}{\includegraphics[angle=-90]{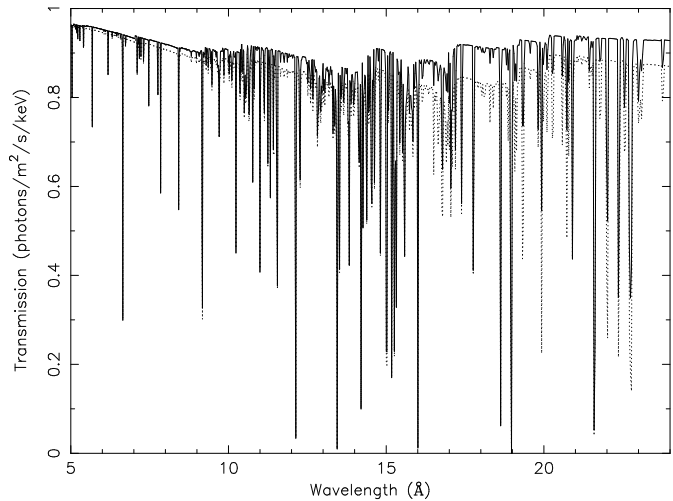}}
\caption{The transmission for the best fit model with three {\it xabs} components for the LETGS (thin solid line) and the HETG (dotted line) spectra. Note that the difference is largest, 10 \% in continuum transmission and 35 \% for the largest difference in absorption lines, for lowly ionized ions which have absorption lines around 20 \AA. Here the MEG effective area is already rather low, and the errors on this lowest ionization parameter are large.} 
\label{fig:shortvar}
\end{figure}

First we studied possible differences in the HETG and LETG spectrum, as the average flux level differed by 30 \% between both observations on a timescale of about 300 ks. In our first approach, we have taken the fluxed MEG spectrum and folded this through the LETGS response matrix. This is possible because of the two times higher spectral resolution of the MEG as compared to the LETGS. We adjusted the MEG continuum to match the higher LETGS continuum. Restricting our analysis to the 1.5--24~\AA\ band, where both instruments have sufficient sensitivity, we find good agreement between both spectra. 
\par
We find no significant difference in the equivalent width of the strongest absorption lines such as the Ly$\alpha$ and Ly$\beta$ lines of \ion{Ne}{x} and \ion{O}{viii} and the resonance lines of \ion{Ne}{ix} and \ion{O}{vii}. There may be a small enhancement of the most highly ionized lines in our spectrum (\ion{Si}{xiv} and \ion{Mg}{xii} Ly$\alpha$ and the \ion{Si}{xiii} resonance line), corresponding to a $\sim$50~\% increase in ionic column density in the LETGS observation. For the \ion{Si}{xiii} line this is uncertain because the line coincides with an instrumental edge in the MEG spectrum. Moreover, at the high ionization parameter where these ions are formed ($\log\xi\sim 2.5$) the opacity of the \ion{O}{viii} Ly$\alpha$ line is even higher than for the Mg and Si lines. Since we do not observe an enhanced column density for \ion{O}{viii}, we conclude that the evidence for enhanced opacity at high ionization parameter during the LETGS observation is uncertain. A formal fit to the difference spectrum shows that any additional high ionization component  emerging during the LETGS observation has a hydrogen column density less than $10^{24}$~m$^{-2}$, or an order of magnitude less than the persistent outflow.
\par
In a different approach, we have fitted the warm absorber in the LETGS spectrum between 1.5 and 24 \AA using a combination of three {\it xabs} components, the results are listed in Table~\ref{tab:comp}. Fitting the HETG spectra with the same ionization parameters and column densities, but fixing the continuum parameters to the best fit parameters (see Table~\ref{tab:cont}) we find an excellent fit. Allowing the ionization and column densities to be free parameters, we obtain the results listed in Table~\ref{tab:comp}. The improvement for allowing the ionization and column densities to be free is $\Delta \chi^2$ = 10 for 4242 degrees of freedom. The maximum difference between the two data sets occurs for the lowest ionization component 3. For the HETG spectra this component has log $\xi$ = $-$0.1 $\pm$ 2.5, this clearly indicates the lack of sensitivity at the longer wavelengths in the HETG spectra. Fig.~\ref{fig:shortvar} shows the transmission for the best fit model for LETGS (solid line) and the HETGS (dotted line). Note that the largest discrepancy is around 20 \AA, where the effective area of the MEG is rather low, and the lines are for the lowest ionized component 3. The flux in the soft band increased by $\sim$50~\% between the HETGS and LETGS observation. Therefore, an increase of $\xi$ by 50~\% is excluded for component 2. However, our data cannot rule out that the ionization parameter responds linearly to continuum flux enhancements for components 1 and 3. We conclude that the difference in ionization parameter of the warm absorber between the HETG and the LETGS observation is small and below the sensitivity in log $\xi$ of $\sim$0.15. 

\begin{table}
\caption{Comparison of the best fit parameters for the LETGS and HETGS spectra between 1.5 and 24 \AA. The labeling of the different components follows Kaastra et al. (\cite{kaastra02}). }
\label{tab:comp}
\begin{tabular}{l|cc|cc}\hline
          & LETGS            & HETGS         & LETGS            & HETGS  \\
          & log $\xi$$^{a}$  & log $\xi$$^{a}$& log N$_{\rm H}$$^{b}$ & log N$_{\rm H}$$^{b}$ \\\hline
1         & 2.26 $\pm$ 0.09  & 2.3 $\pm$ 0.3 & 25.47 $\pm$ 0.06 & 25.36 $\pm$ 0.08  \\
2         & 1.77 $\pm$ 0.04  & 1.9 $\pm$ 0.1 & 24.85 $\pm$ 0.14 & 25.02 $\pm$ 0.12  \\ 
3         & $-$0.2 $\pm$ 0.2 & $-$0.1 $\pm$ 2.5& 24.33 $\pm$ 0.16 & 24.8 $\pm$  0.4 \\
\end{tabular}\\
$^{a}$ Ionization parameter in 10$^{-9}$ W m. \\
$^{b}$ Hydrogen column density in m$^{-2}$.  
\end{table}

A similar situation holds for the time variability during the LETGS observations. Despite the significant flux increase during the observation (see Kaastra et al. \cite{kaastra04}), the signal to noise ratio of our data is insufficient to rule out or confirm a significant response of the warm absorber to the change in ionizing flux.
\par

\subsection{Long term spectral variations}

\begin{figure}
\resizebox{\hsize}{!}{\includegraphics[angle=-90]{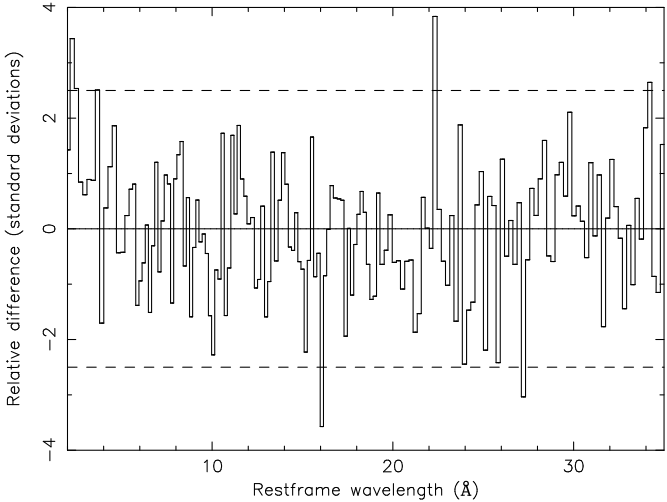}}
\caption{Fit residuals of the 1999 LETGS spectrum minus the fit residuals of the 2002 spectrum, normalized as described in the text. For clarity, the residuals have been rebinned by a factor of 8. The dashed lines indicate the 2.5 $\sigma$ significance level used in the analysis.
}
\label{fig:9902resi}
\end{figure}

We have tested for the presence of long term variations in the warm absorber of
NGC~5548 by comparing our best fit model for the present data with the LETGS
spectrum of December 1999 (Kaastra et al.  \cite{kaastra00}, \cite{kaastra02}).
To do this, we have fit the 1999 spectrum keeping all parameters of the warm absorber and all fluxes of
narrow or broadened emission lines fixed to the values of our 2002 spectrum. Only the parameters of the power law
and modified blackbody component were allowed to vary.  Then we subtracted the relative
fit residuals for our 2002 spectrum from the fit residuals of the 1999 spectrum
(Fig.~\ref{fig:9902resi}). The relative fit residuals $r$ are defined by $r = (o - m)/m$ where $o$ is the observed spectrum, $m$ the best fit model spectrum. Subtracting the fit residuals instead of dividing the spectra allows us to take out continuum variations. Further, as the higher orders are not subtracted any changes in the continuum will produce broad band features in the ratio of two spectra. In Fig.~\ref{fig:9902resi} we normalized the difference of the fit residuals by the standard deviation of this quantity. This erases all remaining errors due to either
shortcomings in the effective area calibration or in the spectral modeling.

\begin{table}
\caption{Features in the difference spectrum of the 1999 and 2002 LETGS
data. We list Gaussian centroids $\lambda$ (in the restframe of 
NGC~5548), $\sigma$ and peak value  $p$; peaks are
expressed as a fraction of the 2002 spectrum at the same energy.}
\label{tab:timdif}
\centerline{
\begin{tabular}{|llr|}
\hline
$\lambda$ (\AA) & $\sigma$ (\AA) & $p$ \\
\hline
16.10$\pm$0.06 & 0.10$^{+0.09}_{-0.04}$ & $-$0.25$\pm$0.09 \\
22.36$\pm$0.02 & 0.03$\pm$0.02          & $+$0.85$\pm$0.29 \\
27.18$\pm$0.07 & 0.08$^{+0.07}_{-0.05}$ & $-$0.27$\pm$0.12 \\
34.13$\pm$0.08 & 0.12$\pm$0.07          & $+$0.34$\pm$0.14 \\
\hline
\end{tabular}
}
\end{table}

The figure shows clear evidence for five features with more than $2.5\sigma$
significance. There is evidence for an increasing flux toward smaller
wavelengths starting around 5~\AA~and resulting in about a 20~\% excess at 2~\AA\
in the 1999 spectrum. This corresponds to a difference in the continuum and not in the warm absorber. The remaining features are not due to continuum variations (Table~\ref{tab:timdif}). As the optical and UV broad emission lines are known to be variable on timescales as short as a day, we identify the broader features with changes in the broad emission lines. We identify the 34.13~\AA\
feature as the red wing of a broad \ion{C}{vi} Ly$\alpha$ line (restframe wavelength
33.74~\AA, hence velocity of $+3500\pm 700$~km~s$^{-1}$). Similarly, the 27.18~\AA\
feature can be identified as the red wing of \ion{C}{vi} Ly$\gamma$ (restframe wavelength 26.99~\AA,
hence velocity of $+2100\pm 800$~km~s$^{-1}$). Apparently, the red wing of the \ion{C}{vi} Ly$\alpha$ line has decreased while the red wing of Ly$\gamma$ increased and no change in the Ly$\beta$ line. However, the process that would cause the Ly$\alpha$ line to decrease while the Ly$\gamma$ line increases is unknown. The intrinsic width, $\sigma$, of the variable parts of both lines corresponds to about 900--1000~km~s$^{-1}$, this is much smaller than the assumed FWHM of 8000 km~s$^{-1}$. The 16.10~\AA\ feature can be identified by a red wing of \ion{O}{viii} Ly$\beta$ (restframe wavelength 16.01~\AA, hence velocity of $+1700\pm 1100$~km~s$^{-1}$). A change in the strength of the broad emission line due to the \ion{O}{vii} triplet can also explain the dip at 21 $-$ 22 \AA.

Finally, the 22.36~\AA\ feature most likely has a different origin.  First, it is
intrinsically narrow and poorly resolved (the upper limit to its Gaussian width
corresponds to 700~km~s$^{-1}$).  Its wavelength coincides within the error bars with
the wavelength of the strongest O~V absorption line (22.374$\pm$0.003~\AA). The 2002 spectrum shows a deep \ion{O}{v} absorption line, which is much weaker in the 1999
spectrum.  This demonstrates that the \ion{O}{v} column is variable on the time scale of a
few years. Note that the \ion{O}{v} line is by far the strongest absorption line in the LETGS spectrum for the ionization parameter log $\xi$ = $-0.2$, the ionization parameter where \ion{O}{v} reaches it maximum concentration. The Fe M-shell lines are formed at the same ionization parameter, but are much weaker.

\section{Discussion\label{sect:disc}}

\subsection{Comparison between the warm absorber models\label{sect:comp}}

In our model B we determined the column density, the average outflow velocity, and the velocity broadening for all observed ions separately (Table~\ref{tab:1v}). The total column densities determined using this method, for the six ions \ion{C}{vi}, \ion{O}{v}, \ion{O}{vii}, \ion{O}{viii}, \ion{Ne}{ix}, and \ion{Si}{xiii} are consistent with the results obtained using method A, with fixed velocity structure. This gives confidence that the column densities as listed in Tables~\ref{tab:1v} and \ref{tab:dnh} are reliable, even if there is substructure to the lines. The average outflow velocity determined for these six ions is consistent with the outflow velocity of the deepest UV components. Comparing our measured column densities with those obtained in the UV for the five outflow velocity components, we conclude that the column density measured for the $-160$~km~s$^{-1}$ component is in fact the blend of UV components 4 and 5. Our $-530$~km~s$^{-1}$ component is a blend of UV components 2 and 3. Components 2 through 5 form one unresolvable blend in X-rays, and only component 1 is clearly separated.

\subsection{Outflow velocity\label{sect:outflow}}
Our method B does not resolve the full velocity structure of the outflow. Therefore one should take care in interpreting the relation between the ionization parameter and the average outflow velocity (Fig.~\ref{fig:belzv}). Namely, all line profiles probably are a blend of five outflow velocities, but due to limited signal to noise ratio and spectral resolution, components 3 and 4 of the UV dominate the blend.
\par
\begin{figure}
 \resizebox{\hsize}{!}{\includegraphics[angle=-90]{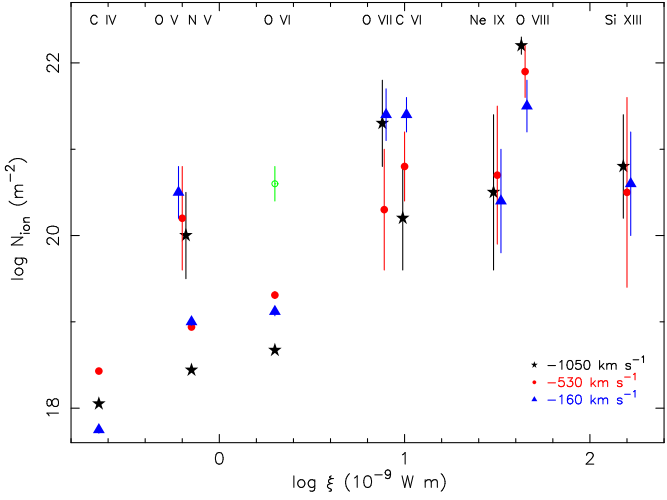}}
\caption{The ionic column density versus the logarithm of the ionization parameter for the three outflow velocity components as measured for six ions in X-rays, 2 ions (\ion{C}{iv} and \ion{N}{v}) measured simultaneously with HST STIS (Crenshaw et al. \cite{crenshaw}) and the lower limits for \ion{O}{vi} from non-simultaneous FUSE data (Brotherton et al. \cite{brotherton}, taking their preferred uncovered model; Arav et al. \cite{arav03}). In calculating the column densities for the UV data we kept the highest outflow value separate, added the two middle ones and the two lower ones (see Sect.~\ref{sect:outflow}). For \ion{O}{vii} the ionization parameter was decreased by 0.05 for easier identification. The open circle indicates the total column density for \ion{O}{vi} obtained from the LETGS and MEG spectra.}
\label{fig:3v}
\end{figure}

In general, there is a trend that more highly ionized ions have a larger average outflow velocity. This could indicate that the $-$1040~km~s$^{-1}$ outflow velocity component becomes more prominent for the more highly ionized ions. The $-$530 km~s$^{-1}$ outflow velocity component dominates for lower ionized ions observed in X-rays and the UV band. To further study this effect we plot the total column density as a function of outflow velocity (Fig.~\ref{fig:3v}) for the six ions of Table~\ref{tab:3v}. We also added \ion{C}{iv} and \ion{N}{v} taken from Crenshaw et al. (\cite{crenshaw}) and \ion{O}{vi} taken from the non-simultaneous FUSE data by Brotherton et al. (\cite{brotherton}). For the UV data we added the two components with the lowest outflow velocity to compare it with the $-$160 km~s$^{-1}$ X-ray component. The two middle outflow velocity components were added to represent the $-$530 km~s$^{-1}$ component. Finally, for \ion{O}{vi} we indicate the total column density as measured in the X-rays. From Fig.~\ref{fig:3v} we note that the ionic column density of the $-$1040 km~s$^{-1}$ velocity component is the smallest of the five velocity components for low $\xi$, while it becomes the largest for highly ionized gas. There is thus a clear difference in ionization structure between the $-$1040 km~s$^{-1}$ outflow component and the lower outflow velocity components. A possible explanation for this difference in ionization structure is that the outflowing wind at $-$1040 km~s$^{-1}$ is less dense, leading to a higher overall ionization of the gas, while overall the column density is smallest.

\subsection{Velocity broadening\label{sect:velb}}

In Table~\ref{tab:3v} we list the measured Gaussian velocity broadening $\sigma$ of the three velocity components as well as the velocity broadening of the UV components 1, $2 - 4$, and 5. For components $2 - 4$ we find a slightly smaller X-ray width, while for component 5 we find a larger width as compared to the UV lines. We attribute this to the blending of (a part of) component 4 into component 5. We note that UV component 4 has $\sigma$ = 62~km~s$^{-1}$ compared to 68~km~s$^{-1}$ in component 3 and only 18~km~s$^{-1}$ in component 2, hence we need approximately half of component 4 to blend into component 5 in order to explain the difference. 
\par
Interestingly, we find a much smaller velocity broadening for UV component 1, the only one we can resolve. Modeling the \ion{O}{viii} Ly$\alpha$ line with a column density of 10$^{22.2}$ m$^{-2}$ and a velocity dispersion $\sigma$ of 40~km~s$^{-1}$, we find that the line is heavily saturated, and effectively produces a FWHM of 240~km~s$^{-1}$. This is similar to the 222 $\pm$ 18~km~s$^{-1}$ value measured from the UV spectrum. We are, however, able to obtain the velocity broadening $\sigma$ from the line ratios of the non-saturated Lyman series of \ion{O}{viii} and the other five ions.
\par
In Table~\ref{tab:1v} the velocity broadening $\sigma$ is listed based upon model B. It is small compared to the velocity range of the components measured from the UV spectra. This indicates that we do not fully resolve the blend of the five outflow velocity components. The velocity broadening in our case is mainly determined from equivalent width ratios of absorption lines from the same ion. The velocity broadening for the majority of absorption lines is less than 220~km~s$^{-1}$. This is consistent with detecting mainly the $-$667~km~s$^{-1}$, $-$530~km~s$^{-1}$ and $-$336~km~s$^{-1}$ outflow components, which are the dominant velocity components in the UV. These three velocity components form an unresolved blend in the X-rays. Plotting the velocity broadening versus the ionization parameter produces a scatter plot. 

\section{Ionization structure\label{sect:xi}}
An important question about the physical nature of these outflows is whether they occur as clumps in an outflow or in a more homogeneous wind. In the first case these clumps should be in pressure equilibrium with the outflowing less dense wind, with a finite number of solutions for temperature and ionization $\xi$ for the same constant pressure ionization $\Xi$ value. In the second case one expects a continuous ionization distribution, which depends on the density of the particular part of the wind. Thus less dense material will be higher ionized, while more dense regions will be lower ionized. In this section we elaborate on both scenario's, confronting them with our observations.

\subsection{Clumps in a wind\label{sect:mod}}
In several AGN outflow scenario's, instabilities in the wind may cause the formation of clumps. Such clumps can survive for longer times if they are in pressure equilibrium with the surrounding wind and are on the stable branch of the $T(\Xi)$ curve. Where the ionization parameter for constant pressure, $\Xi$, is given by $\Xi = L/(4\pi c r^2 P)$ = 0.961 $\times$ 10$^{4}$ $\xi/T$, with $L$ the luminosity, $P$ the pressure, $r$ the distance from the ionizing source, c the speed of light and $T$ the temperature. Ionization codes, like XSTAR (Kallman \& Krolik \cite{kallman}) predict a specific ionization range for which pressure equilibrium can occur (Krolik \& Kriss \cite{krolik}). Krongold et al. (\cite{krongold}) claim that there are only two ionization components in such an equilibrium in the Seyfert 1 galaxy NGC~3783. Netzer et al. (\cite{netzer}) needed three ionization components in pressure equilibrium to fit the same NGC~3783 spectrum. Ogle et al. (\cite{ogle}) find that in the case of NGC 4051 the different ionization components are not in pressure equilibrium. Another possibility is that the clumps are magnetically confined.
\par
In order to test the presence of a finite number of ionization components, each with its own value for $\xi$ and $N_{\rm H}$, we fitted the ionic column densities obtained with model B (Table~\ref{tab:dnh}) to a model with a finite number of ionization components. Since there are potential problems with the iron ionization balance (see Sect.~\ref{sect:modelc}), we did our analysis separately for iron and the other elements. The results of these fits are listed in Table~\ref{tab:10} and Table~\ref{tab:11}. 
\par
Using the measured column densities for all ions, except iron, we notice that the fit does not improve if more than three ionization components are fitted. The program even prefers to split up one component rather than add another in the case we fitted for five ionization components. The fit is never statistically acceptable, which is unlikely due to abundance effects as most elements have abundances consistent with solar. As an extra test we decided to fit oxygen, silicon and sulfur separately. However, due to the smaller span in ionization range and the fewer points (maximum of five), all were well fit with three ionization components.

\begin{table}
\caption{Fits with a finite number of ionization components to the column densities of all elements but iron derived with model B, assuming solar abundances. The number of ionization components is indicated by N. In the last column we give the significance according to an F-test of the added component.}
\begin{center}
\begin{tabular}{lllll}\hline
\label{tab:10}
N & $\chi^2$/d.o.f. & log $\xi$ & log $N_{\rm H}$  & sign. \\
         &          & 10$^{-9}$ W m & m$^{-2}$     &       \\\hline
1        & 147/22   & 1.2       &  25.5            &       \\
2        & 86/20    & 2.45, 1.15 &  26.1, 25.4     & 89  \\
3        & 83/18    & 2.47, 1.17, $-$0.70 & 26.1, 25.4, 23.6 & 53\\
\end{tabular}
\end{center}
\end{table}

\begin{table}
\caption{Fits with a finite number of ionization components to the iron column densities for the RGS data (Steenbrugge et al. \cite{steen}). The number of ionization components is indicated by N. In the last column we give the significance according to an F-test of the added component.}
\begin{center}
\begin{tabular}{lllll}\hline
\label{tab:11}
N & $\chi^2$/d.o.f. & log $\xi$ & log $N_{\rm H}$ & sign.  \\
  &                 & 10$^{-9}$ W m & m$^{-2}$    &        \\\hline
1 & 79/15      & 0.22      & 25.2                 &        \\
2 & 22.9/13    & 2.50, 0.20 & 25.7, 25.2          & 98.8   \\
3 & 10.5/11    & 2.57, 1.61, 0.20 & 25.7, 25.1, 25.2 & 90.9  \\
4 & 7.2/9      & 2.57, 1.62, 0.65 & 25.7, 25.1, 25.0 & 72.7 \\
  &            & $-$0.03          & 25.0             &      \\
5 & 4.9/7      & 3.10, 2.48, 1.56 & 25.8, 25.6, 25.1 &  71.1\\
  &            & 0.65, $-$0.03    & 25.0, 25.0    &    \\
\end{tabular}
\end{center}
\end{table}

\begin{figure}
 \resizebox{\hsize}{!}{\includegraphics[angle=-90]{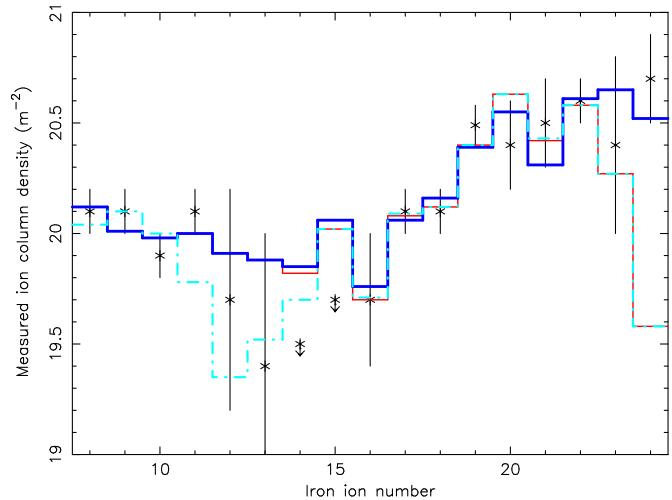}}
\caption{The column densities for iron as measured by the RGS instrument, with the fit for a model with four (dash-dot line) and five (thick solid line) ionization components. The fit with four ionization components clearly underpredict the \ion{Fe}{xxiv} column density.}
\label{fig:rgs_fe4}
\end{figure}

For iron we decided to use the column densities measured by the RGS (Steenbrugge et al. \cite{steen}), as here several more ions have a measured column density and the errors are somewhat smaller. For \ion{Fe}{xii} and \ion{Fe}{xxiv}, which have the same measured column density in the RGS and {\it Chandra} spectra, we used the smaller {\it Chandra} error bars. To fit these column densities adequately, one needs at least five ionization components (log $\xi$ = 3.10, 2.48, 1.56, 0.65, $-$0.03). In Fig.~\ref{fig:rgs_fe4} we show the fit to the measured column densities with four and five ionization components. The main problem with the fit for four ionization components is \ion{Fe}{xxiv}, which is severely under-predicted. As the RGS and {\it Chandra} spectra measure the same column density for this ion, this is a rather certain measurement. 

\begin{figure}
 \resizebox{\hsize}{!}{\includegraphics[angle=-90]{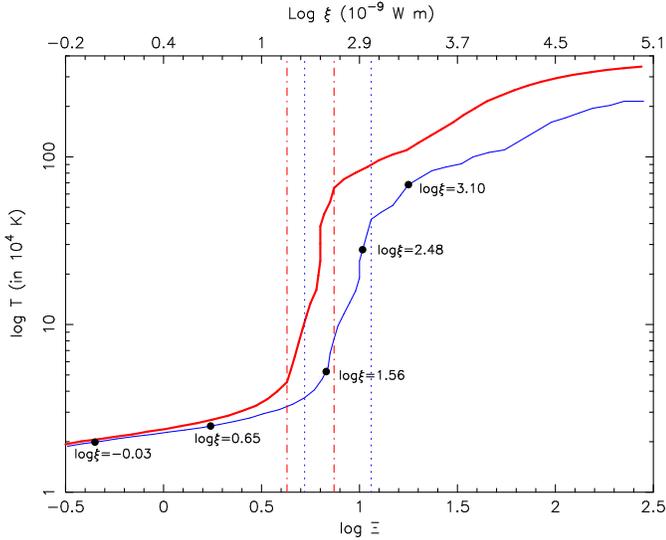}}
\caption{The temperature versus ionization parameter for constant pressure $\Xi$ diagram for the SED used in the present analysis (thick line) and for the SED used in earlier LETGS (Kaastra et al. \cite{kaastra02}) and RGS (Steenbrugge et al. \cite{steen}) analysis (thin line). The $\xi$ values corresponding to the currently used SED are indicated on the x-axis on top. The five components derived from fitting iron in the RGS spectrum are indicated. The dotted lines indicate the boundaries of the marginally stable branch for the SED used in analyzing the RGS data; the dash-dotted lines are the same boundaries, but for the current used SED.}
\label{fig:taver}
\end{figure}

We conclude that we need at least five different ionization components to explain the measured ionic column densities in NGC~5548. These components span a wide range in ionization parameter (log $\xi$ between $-0.03$ and 3.10). Are these ionization components in pressure equilibrium? In Fig.~\ref{fig:taver} the ionization parameter if pressure is held constant ($\Xi$) versus temperature is shown for the SED used in the current analysis and the earlier RGS and LETGS analysis. The marginally stable part of the curve indicates a constant $nT$ value, therefore components with a different $\xi$ are in pressure equilibrium on the marginally stable branch from log $\xi$ = 1.3 to 2.7. The curve has nowhere a negative slope, the smallest positive slope is between log $\xi = 1.4 - 1.5$. It should be noted that the difference between the results produced by different codes can be large, but substituting CLOUDY 95.6 (Ferland \cite{ferland}) for XSTAR we find that still not all components are in pressure equilibrium.
\par
For the three different ionization components as observed in the earlier LETGS and RGS spectra of NGC~5548 (Kaastra et al. \cite{kaastra02}; Steenbrugge et al. \cite{steen}), the lowest ionization component has log $\xi$ = 0.4, which cannot be in equilibrium with the other two components at log $\xi$ = 1.98 and 2.67. If we take the five ionization components from the iron analysis, then the highest and the lowest ionization component cannot be in pressure equilibrium. This clearly suggests that the current photo-ionization balance models are either too simple, or that at least the lowest ionization component is confined by another process, possibly magnetic. As the lowest ionization component does have the same kinematics as the higher ionized absorbers, all absorbers must co-exist in a single confined outflow. The result that not all ionization components are in pressure equilibrium is rather robust, even accounting for inaccuracies in the photoionization balance codes, in particular for lowly ionized iron, or the inclusion of the blue bump.

Another possible method to distinguish between a continuous and discrete ionization is to look for spectral variability. No spectral variability of the warm absorber, with the exception of \ion{O}{v}, was detected although the source was observed at a $2 - 10$ keV luminosity of 4.9 $\times$ 10$^{36}$ W (1999 LETGS, Kaastra et al. \cite{kaastra02}), 5.7 $\times$ 10$^{36}$ W (2001 RGS, Steenbrugge et al. \cite{steen}) and 4.0 $\times$ 10$^{36}$ W (present LETGS spectrum). However, if the ionization parameter varies linearly with luminosity, then no variation in ionization is expected to be detectable. Further complicating the possible detection of changes in the ionization parameters is the fact that in the UV the different velocity components show opposite column density variations for the same ion. The net effect will be hard to discern in the X-rays for velocity components 2 $-$ 5. Component 1, which can be resolved with {\it Chandra} is the most variable component, and should be studied in the future for signs of variability. A further lack of variation in the ionization structure of component 1 possibly indicates that the ionization distribution is continuous. In the case of two or three ionization components, a luminosity change should result in a shift in the ionization parameters. In a continuous model one would expect a decrease in column density for the lowest ionization states and an increase in highly ionized material for an increase in luminosity. However, as the warm absorber is detected up to \ion{Fe}{xxiv}, and our sensitivity for \ion{Fe}{xxv} and \ion{Fe}{xxvi} is insufficient, this is hard to measure with this data set. However, for the long NGC~3783 XMM-{\it Newton} observation this behaviour was indeed detected. Behar et al. (\cite{behar}) studying the RGS spectra detected no spectral variability in the lowly ionized absorber, and in particular no change in the iron UTA. Reeves et al. (\cite{reeves}) however did detect spectral variability of the highly ionized absorber with the EPIC pn instrument. 

\subsection{Continuous ionization distribution\label{sect:conion}}

\begin{figure}
 \resizebox{\hsize}{!}{\includegraphics[angle=-90]{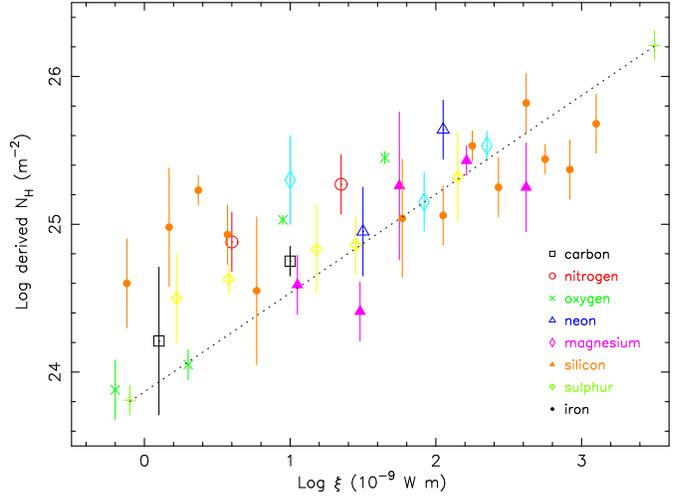}}
\caption{The total hydrogen column density $N_{\rm H}$, assuming solar abundances (Anders \& Grevesse \cite{anders}) derived using Eq.~\ref{eq:n}, plotted versus ionization parameter. The ionic column densities were taken from Table~\ref{tab:1v} and Table~\ref{tab:dnh} assuming a velocity broadening of 140~km~s$^{-1}$. For clarity, no upper limits have been plotted. The best fit results for model D are plotted as the two crosses connected by a dotted line.}
\label{fig:xi-nh}
\end{figure}

We decided to derive the total hydrogen column density as a function of $\xi$ for each ion, similar to Steenbrugge et al. (\cite{steen}). This in contrast to the {\it warm} model where we determined the total column density at the two extrema. Contrary to the previous analysis we did not assume that the ion column densities are predominantly determined by the ionization parameter at which the relative concentration is at its maximum. In the present analysis we took into account that each ion is formed over a range of ionization parameter similar to the {\it warm} model. We implemented this by assuming a power-law distribution, as this function well describes the RGS data (Steenbrugge et al. \cite{steen}), and appears to be sufficient in the {\it warm} model:

\begin{equation}
\label{eq:xi}
\xi {\rm d}N_{\rm H} (\xi)/ {\rm d}\xi = A \xi^{\alpha}
\end{equation}

with a given normalization $A$ and index $\alpha$. We calculate the predicted ionic column densities $N_{\rm i,p}$ for this distribution similar to the {\it warm} model. As the ionization parameter range in reality will be finite, this power law needs to have a cut-off value, which was chosen to be log $\xi$ = 4. We will use the results from model B to determine $A$ and $\alpha$. If $f_{\rm i} (\xi)$ is the ionic concentration relative to hydrogen at a given ionization parameter $\xi$, then $N_{\rm i,p}$ is given by $N_{\rm i,p} = \int_0^\infty f_{\rm i} (\xi) A \xi^{\alpha-1}$ d$\xi$. For each ion the effective ionization parameter $\xi_{\rm i}$ is defined as

\begin{equation}
\label{eq:m}
\xi_{\rm i} \equiv \frac {\int_0^\infty f_{\rm i}(\xi) \xi {\rm d}N_{\rm H}(\xi)} {\int_0^\infty f_{\rm i}(\xi) {\rm d}N_{\rm H}(\xi)} = \frac {\int_0^\infty f_{\rm i} (\xi) \xi^{\alpha} {\rm d}\xi}{\int_0^\infty f_{\rm i}(\xi) \xi^{\alpha-1} {\rm d}\xi}
\end{equation}

This value of $\xi_{\rm i}$ is typically 25 \% larger than the value of $\xi$ for which $f_{\rm i} (\xi)$ reaches its maximum. This increase is solely due to our assumption of a power-law distribution. This can be used to derive the {\it equivalent hydrogen column density $N_{\rm H,i}$ for each ion} (not to be confused with the integrated column density over all $\xi$ values) from the observed ionic column densities $N_{\rm i,obs}$, since obviously we should have $N_{\rm i,obs}$ = $N_{\rm i,p}$. This yields equation~\ref{eq:n}: 

\begin{equation}
\label{eq:n}
N_{\rm H,i} \equiv \xi_{\rm i} \frac {{\rm d}N_{\rm H}(\xi_{\rm i})}{{\rm d}\xi} = \frac {N_{\rm i,obs}}{N_{\rm i,p}} A \xi_{\rm i}^{\alpha}
\end{equation}

The calculated total hydrogen column density using this method, instead of assuming that the ion is formed at the ionization parameter where $f_{\rm i} (\xi)$ has its maximum, is larger by at most a factor of 3. Ion concentrations were determined using a set of runs with XSTAR (Kallman \& Krolik \cite{kallman}) for a thin photoionized layer with solar abundances (Anders \& Grevesse \cite{anders}). 
\par
The derived hydrogen column densities $N_{\rm H,i}$ versus the ionization parameter $\xi_{\rm i}$ are presented in Fig.~\ref{fig:xi-nh}. For lower ionized ions (log $\xi$ $<$ $- 0.7$) we can only determine upper limits to the column density, in contrast to the earlier RGS results (Steenbrugge et al. \cite{steen}). This results from the lower effective area of the LETGS compared to the RGS in the iron UTA region. However, due to the longer wavelength band, we do detect many previously unobservable medium ionized ions. This strengthens the relationship between the hydrogen column density and ionization parameter in the range log $\xi$ = 0 to 1. The column density increases by nearly 2 decades for a 3 decades increase in $\xi$, consistent with Steenbrugge et al. (\cite{steen}).    
\par
A formal fit of the column densities of all ions except iron yields A = 10$^{24.4 \pm 0.1}$ (at log $\xi$ = 0) and $\alpha$ = 0.40 $\pm$ 0.05 with $\chi^2$ = 107 for 30 d.o.f. A similar fit to the iron ions only yields A = 10$^{25.0 \pm 0.2}$ (at log $\xi$ = 0) and $\alpha$ = 0.2 $\pm$ 0.1, with a $\chi^2$ = 16 for 13 d.o.f. 
\par
The relationship obtained is rather tight, and the assumption of solar abundances seems justified, if we do not include those ions which may suffer from systematic bias in their column density determinations (Sect~\ref{sect:id}). If the same analysis is repeated for sodium, aluminum, argon and nickel, we find that these elements are overabundant compared to solar, except for sodium. This strengthens our conclusion that these ions may suffer from strong systematic errors in their column density estimates. These ions are therefore not plotted in Fig.~\ref{fig:xi-nh}. 

\section{The outflow}
\subsection{Outflow geometry\label{sect:geom}}
In trying to get a physical picture of the outflows in AGN it is important that we know the geometry, as well as the ionization structure. By estimating the opening angle of the outflow, we can distinguish between a spherical outflow model or a model with localized streams. The model by Elvis (\cite{elvis}) prefers a localized stream. The model by Urry \& Padovani (\cite{urry}), on the other hand predicts a less collimated outflow. 
\par
Similar to Steenbrugge et al. (\cite{steen}) we can obtain an upper limit to the opening angle for the outflow (eq.~\ref{eq:1}). We use the mass conservation formula ($\dot{M}_{\rm loss}$ = $\Omega m_{\rm p} n_e r_0^2 v$), with $r_0$ the distance from the ionizing source, $n_e$ the electron density, $m_{\rm p}$ the proton mass, $v$ the outflow velocity and $\Omega$ the opening angle. Further we assume that the system is stationary and that the mass loss rate is equal or less than the mass accretion rate $\dot{M}_{\rm acc} = L/\eta c^2$, with L the bolometric luminosity and $\eta$ the accretion efficiency. The outflow velocities of the wind are taken from the UV measurement. This leads to the following constraint on the solid angle $\Omega$:
\begin{equation}
\label{eq:1}
\Omega \leq \frac{\xi}{\eta c^2 m_{\rm p} v}  
\end{equation}
where we used: 
\begin{equation}
\label{eq:l}
\xi = \frac{L}{n_e r^2} 
\end{equation}
\par
We assume a Schwarzschild black hole, with an accretion efficiency $\eta$ = 0.057 (Thorne \cite{thorne}). For a Kerr black hole, with an efficiency of 0.30 (Thorne \cite{thorne}), the opening angle is reduced by a factor of 5.4. Even higher efficiencies will further reduce the opening angle. Assuming that only 50 \% of the mass that is accreted is lost through an outflow halves the upper limit to the opening angle. 
\par
As the ionization parameter, as well as the outflow velocities measured in this analysis closely resemble those found with the RGS, also the resulting upper limits for $\Omega$ have a similar range (but now corrected for the extra $4 \pi$). Table~\ref{tab:angle} lists the upper limits to the opening angle for the five different outflow velocity components observed in the UV and the ionization range observed in the X-rays. 

\begin{table}
\caption{The upper limit to the opening angle as calculated from eq.~\ref{eq:1} for the five outflow velocities measured in the UV, over the ionization range observed in the {\it Chandra} spectra. The angles are given in sr.}
\label{tab:angle}
\begin{tabular}{llllll}\\
           & 5       & 4       & 3       & 2        & 1       \\\hline
log $\xi^a$ & $-$166  & $-$336  & $-$530  & $-$667  & $-$1041 \\
            & km s$^{-1}$&km s$^{-1}$&km s$^{-1}$&km s$^{-1}$&km s$^{-1}$ \\\hline
0    & 7$\times$10$^{-4}$& 3.5$\times$10$^{-4}$ & 2$\times$10$^{-4}$ & 2$\times$10$^{-4}$ & 1$\times$10$^{-4}$ \\
1    & 7$\times$10$^{-3}$& 3.5$\times$10$^{-3}$ & 2$\times$10$^{-3}$ & 2$\times$10$^{-3}$ & 1$\times$10$^{-3}$ \\
2    & 7$\times$10$^{-2}$& 3.5$\times$10$^{-2}$ & 2$\times$10$^{-2}$ & 2$\times$10$^{-2}$ & 1$\times$10$^{-2}$ \\
3    & 7$\times$10$^{-1}$& 3.5$\times$10$^{-1}$ & 2$\times$10$^{-1}$ & 2$\times$10$^{-1}$ & 1$\times$10$^{-1}$ \\ 
\end{tabular}\\
$^{a}$ in 10$^{-9}$ W m.
\end{table}

We thus find that the opening angle of the outflow is very small, down to 10$^{-4}$ sr for the lowest ionization component. This implies that the wind occurs in the form of narrow streamers. The range in ionization parameters at a given outflow velocity component then implies density stratification perpendicular to the outflow.
\par
A further quantitative diagnostic of the thickness of the outflow can be derived from Fig.~\ref{fig:xi-nh}. There is a clear power-law relation between the total hydrogen column density and the ionization parameter. This power-law relation is also established by the fact that one can fit the spectra with the {\it warm} model, using only two points, the lowest and highest ionization state observed. From Fig.~\ref{fig:xi-nh} we find a power-law slope $\alpha$ = 0.40 $\pm$ 0.05, and a normalization log A = 24.4 $\pm$ 0.1 at log $\xi$ = 0 (see Sect.~\ref{sect:conion}). Similar to Steenbrugge et al. (\cite{steen}) we combine Eq.~\ref{eq:xi} with Eq.~\ref{eq:l} and use d$N_{\rm H}$ = $n_e$ d$r$ with $r$ the distance from the axis of the streamer, resulting in a simple differential equation for d$\xi$/d$r$ with formal solution $n/n_o$ = (1 + $r/r_o$)$^{-\beta}$ where $n_o$ is the density at the axis and $r_o$ = $A \xi_o^{\alpha + 1}$/$n_o (\alpha + 1)$ and $\beta$ = 1/($\alpha$ + 1). As we do not know the distance $r_o$ of the absorber to the nucleus of the AGN, we do not know $n_o$; however we know $r_o n_o$ = 1.0 $\times$ 10$^{24}$ m$^{-2}$ at $\xi$ = 0 and $\beta$ = 0.71 $\pm$ 0.01. Accordingly, at large distance from the axis of the streamer (low density, high ionization parameter), the density scales as $n \sim r^{-0.71}$. Steenbrugge et al. (\cite{steen}) found a similar result, but due to the detection and overabundance of the iron UTA with the RGS, the power-law slope was less well determined.

\subsection{Mass loss through outflow}
Outflows from AGN are potentially important contributors to the enrichment of the IGM. A critical factor is the amount of mass that escapes from the host galaxy through the observed outflows. Here we estimate an upper limit to the mass loss and discuss briefly the difficulties associated with this estimate.
\par
First the outflowing gas has to escape the potential of the massive black hole. Matter can escape if its radial velocity $v_r$ is larger than the escape velocity $v_e=\sqrt{2GM/r_0}$. Inserting a black hole mass of $6.8\times 10^{7}$~M$_{\odot}$ (Wandel \cite{wandel}) we find that matter escapes only if $r_0 > (0.58/v_r^2)$~pc with $v_r$ expressed in units of 1000~km/s. The measured outflow velocities are only lower limits, namely the velocity component along our line of sight, and range between $-$1041~km\,s$^{-1}$ and $-$166~km\,s$^{-1}$. Taking the measured outflow velocities, then $r_0$ needed for the outflow to escape corresponds to 0.5--20~pc. The location of the wind is thus important in determining whether the gas will escape or not. Crenshaw et al. (2003) argue that the wind should cover at least a part of the inner, high ionization narrow line region (NLR), while Arav et al. (2002) argue that the narrow line region is not covered. The NLR spans at least the distance range of 1~pc (highly ionized) to 70~pc (lowly ionized) (Kraemer et al. \cite{kraemer98}). If the argument of Crenshaw et al. (2003) holds, then in particular the highest velocity gas may be able to escape. If the $v_r$ is significantly higher than the measured $v$, the lower velocity components will also escape.
\par
After escaping the immediate environment of the central black hole, the gas must escape from the bulge and halo of the galaxy. The stellar velocity dispersion of the bulge is 180~km\,s$^{-1}$ (Ferrarese et al. \cite{ferrarese}), and hence it is likely that most of the gas can escape the bulge provided it has not been decelerated. The effects of the disk and halo of NGC~5548 are expected to be of equal or less importance.
\par
We neglect the possibility that ram pressure in the host galaxy terminates the outflow. Wilson \& Ulvestad (\cite{wilson}) note the possibility that the linear radio structure observed in the host galaxy of NGC~5548 is due to material ejected from the nucleus and stopped due to ram pressure. This seems to indicate that the outflows as observed in the UV and the X-rays can escape the nucleus, whether or not the host galaxy.
\par
If all gas escapes, then our earlier argument that the total mass loss through the wind must be smaller than the accretion rate through the disk leads to an upper limit of the mass loss of about 0.3~M$_{\odot}$\,yr$^{-1}$. NGC~5548 had a major interaction with another galaxy some 0.6--1.0~Gyr ago (Tyson et al. \cite{tyson}); assuming that the AGN phase lasts for at least 0.6~Gyr with a steady mass loss rate, the upper limit to the mass enrichment of the IGM is $2\times 10^8$~M$_{\odot}$.

\section{Summary and conclusions}
Our long Chandra observation of NGC~5548 allowed us to obtain a unique, high signal to noise, high spectral resolution, broadband spectrum of this Seyfert~1 galaxy.
\par
The fluxes measured for the Fe-K line and the narrow emission lines are within 1 $\sigma$ consistent with earlier {\it Chandra} and XMM-{\it Newton} spectra. There is thus no evidence for long-term variability. The only narrow feature that shows column density variations is \ion{O}{v}.
\par
We found clear evidence for broad emission lines, similar to the optical and UV Broad Line Region lines. In particular \ion{C}{vi}, \ion{O}{vii} and \ion{O}{viii} showed significant broad lines. From a comparison of our spectrum taken in 2002 with the earlier LETGS observation of 1999, we find significant changes in the red wing (at 1700-3500~km~s$^{-1}$) of \ion{C}{vi} lines and the \ion{O}{viii} Ly$\beta$ line. A more extensive discussion of these broad lines will be given in a forthcoming paper (Steenbrugge et al. 2004). The presence of these broad lines affects our estimates of column densities in the warm wind, by factors up to 3, a similar situation as has been found in the UV band (Arav et al. 2002).
\par
We find evidence for the presence of an inner-shell X-ray absorption line of \ion{N}{v} at 29.42~\AA. The long exposure time and the combination of LETGS and HETGS spectra allows us to fit the warm absorber using three different spectral models. Based upon these models, we conclude that there is a good agreement between the properties of the outflow as measured through UV absorption lines and X-ray absorption lines, although in the X-ray band we do not fully resolve the spectral lines. But line centroids and derived line width are in good agreement. We find that the highest velocity component 1 at $-$1040~km~s$^{-1}$ has a different ionization structure than the other components 2--5. 
\par
We have compared our results with models for the ionization structure of a 
wind. We conclude that our data are not in agreement with a model with 
discrete clumps that are in gas pressure equilibrium with the surrounding medium. 
This conclusion is based upon the following arguments: 
\begin{enumerate}
\item
the need for at least five discrete ionization components to span the broad range (at least 3 orders of magnitude) of ionization parameter $\xi$; 
\item
we find, assuming our SED, a limited range in $\xi$ for which multiple solutions of constant $\Xi$ can co-exist. The lowest ionization parameter determined from the data is incompatible with this range;
\item
the fact that there are five discrete velocity components, instead of a continuous velocity range. At least the $-$1040~km~s$^{-1}$ component spans the observed range in ionization.
\end{enumerate}

Our data are in agreement with a model with a continuous distribution
of $N_{\mathrm H}$ as a function of $\xi$. Comparing the mass outflow through 
the wind with the accretion rate onto the black hole, we derived upper limits 
to the solid angle sustained by the outflowing wind. These upper limits are 
of the order of $10^{-4}$~sr for the lowest ionized gas in velocity component 
1. We conclude that the wind occurs in the form of narrow streams with density variations perpendicular to the flow velocity. Most likely these streamers are caused by accretion disk 
instabilities and cross our line of sight when being radiatively accelerated. 
Assuming density stratification, the dependency of $N_{\mathrm H}$ upon $\xi$ gives the density profile
across these streamers. At large distances the density is proportional to 
$r^{-0.71}$, with $r$ the distance from the densest part of the streamer.

Unfortunately the LETGS is not sensitive enough to detect significant 
variations of the warm absorber as a response to the continuum flare 
occurring during the LETGS observation. Reverberation studies with more 
sensitive X-ray instruments will allow us to determine the densities and 
hence the location of the wind. However regular monitoring of NGC~5548 in 
both X-rays and UV will allow us to derive important constraints on the long 
term variability of the wind. A clear example is the variability of 
\ion{O}{v} between 1999 and 2002 that we found with our observations.

\section*{ACKNOWLEDGMENTS}
 
SRON National Institute for Space Research is supported financially by NWO, the Netherlands Organization for Scientific Research. We thank the referees for their constructive comments.

\noindent{\bf APPENDIX A}
\nopagebreak
\samepage
\begin{table*}[b!]
\begin{center}
{\footnotesize
\caption{ The best fit column densities (in m$^{-2}$) as measured using model B. The outflow velocity (in km~s$^{-1}$) was taken from Table~\ref{tab:1v} or a function of ionization and frozen. The velocity broadening (in km~s$^{-1}$) was also frozen during the fit. In the last two columns we list the ionization parameter (in 10$^{-9}$ W m) for which the ion has its maximum column density. The column densities quoted are for those measured in the fit including the broad emission lines (Sect.~\ref{sect:bel}). All ions with uncertain column densities are indicated by * (see Sect.~\ref{sect:id}).}
\vspace*{-0.6cm}
\label{tab:dnh}
\begin{tabular}{lllrrrllllrr}\hline
ion & log $N_{\rm ion}$ & log $N_{\rm ion}$ & $v$ & log $\xi$ & log $U$ & ion & log $N_{\rm ion}$ & log $N_{\rm ion}$ & $v$ & log $\xi$  & log $U$ \\
& $\sigma$ = 140 & $\sigma$ = 70 &   &   &   &   & $\sigma$ = 140 & $\sigma$ = 70 &   &   &   \\\hline

\ion{C}{iv}$^{*}$& 21.3 $\pm$ 0.1 & 20.6 $\pm$ 0.4 & 500 & $-$0.65 & $-$2.25 & \ion{Ar}{xiii}$^{*}$ & 19.6 $\pm$ 0.6 & 19.5 $\pm$ 0.5 & 700 & 1.5     & $-$0.10 \\
\ion{C}{v}     & 20.6 $\pm$ 0.2 & 21.2 $\pm$ 0.1 & 530 & 0.1     & $-$1.50 & \ion{Ar}{xiv}$^{*}$  & $<$ 20.0      & $<$ 19.5         & 700   & 1.75  & 0.15      \\
\ion{C}{vi}    & 21.2 $\pm$ 0.1 & 21.5 $\pm$ 0.1 & 480  & 1.0    & $-$0.60 & \ion{Ar}{xv}$^{*}$   & $<$ 20.0      & $<$ 20.1         & 700   & 1.95  & 0.35      \\
\ion{N}{v}   & $<$ 19.8 & $<$ 19.8  & $-$ & $-$0.15 & $-$1.75              & \ion{Ar}{xvi}$^{*}$  & $<$ 21.7      & 22.1 $\pm$ 0.1  & 800  & 2.30  & 0.70      \\
\ion{N}{vi}    & 20.3 $\pm$ 0.2 & 21.2 $\pm$ 0.2 & 550 & 0.5     & $-$1.10 & \ion{Ar}{xvii}$^{*}$ & $<$ 20.0      & $<$ 20.2        & 800  & 2.60  & 1.00     \\
\ion{N}{vii}   & 21.2 $\pm$ 0.2  & 21.7 $\pm$ 0.1 & 320 & 1.35   & $-$0.25 & \ion{Ar}{xviii}$^{*}$& $<$ 21.2      & $<$ 20.7        & 800  & 3.05  & 1.45      \\
\ion{O}{iii}$^{*}$   & $<$ 20.5 & 20.2 $\pm$ 0.4 & 400 & $-$1.7  & $-$3.30 & \ion{Ca}{xi}$^{*}$   & $<$ 18.8      & 19.2 $\pm$ 0.4  & 600  & 0.70  & $-$0.90   \\
\ion{O}{iv}$^{*}$& 20.3 $\pm$ 0.3 & 20.2 $\pm$ 0.3 & 400 & $-$0.85 & $-$2.45 & \ion{Ca}{xii}$^{*}$  & $<$ 19.4      & $<$ 19.5        & 600  & 1.00  & $-$0.60   \\
\ion{O}{v}     & 20.6 $\pm$ 0.2 & 20.9 $\pm$ 0.2 & 530 & $-$0.2  & $-$1.80 & \ion{Ca}{xiii}$^{*}$ & $<$ 19.6      & 19.3 $\pm$ 0.8  & 700  & 1.40  & $-$0.20  \\
\ion{O}{vi}    & 20.4 $\pm$ 0.2 & 21.0 $\pm$ 0.1 & 380 & 0.3     & $-$1.30 & \ion{Ca}{xiv}$^{*}$  & 19.7 $\pm$ 0.3& 19.6 $\pm$ 0.3  & 700  & 1.60  & 0.00      \\
\ion{O}{vii}   & 21.83 $\pm$ 0.05& 21.94 $\pm$ 0.03& 590 & 0.95  & $-$0.65 & \ion{Ca}{xv}$^{*}$   & $<$ 18.6      & $<$ 19.2        & 700  & 2.00  & 0.40       \\
\ion{O}{viii}  & 22.31 $\pm$ 0.04& 22.37 $\pm$ 0.03& 540 & 1.65    & 0.05  & \ion{Ca}{xvi}$^{*}$  & $<$ 20.2      & $<$ 19.8        & 800  & 2.25  & 0.65      \\
\ion{Ne}{vii}$^{*}$  & 21.6 $\pm$ 0.1 & $<$ 20.7       & 500 & 0.35    & $-$1.25 & \ion{Ca}{xvii}$^{*}$ & 20.0 $\pm$ 0.5& $<$ 19.8        & 800  & 2.40  & 0.80      \\
\ion{Ne}{viii}$^{*}$ & 21.4 $\pm$ 0.1 & 20.3 $\pm$ 0.8 & 500 & 0.9     & $-$0.70 & \ion{Ca}{xviii}$^{*}$& 20.2 $\pm$ 0.4& $<$ 20.2        & 800  & 2.60  & 1.00      \\
\ion{Ne}{ix}   & 21.3 $\pm$ 0.1 & 20.9 $\pm$ 0.2 & 660 & 1.5     & $-$0.10 & \ion{Ca}{xix}$^{*}$  & $<$ 20.9      & $<$ 20.5        & 800  & 2.80  & 1.20      \\
\ion{Ne}{x}    & 21.5 $\pm$ 0.2 & 21.3 $\pm$ 0.1 & 830 & 2.05    & 0.45    & \ion{Ca}{xx}$^{*}$   & $<$ 20.9      & $<$ 20.4        & 800  & 3.25  & 1.65     \\
\ion{Na}{vii}$^{*}$  & 21.4 $\pm$ 0.1 & $<$ 20.3 & 500 & 0.25    & $-$1.35 & \ion{Fe}{iii}$^{*}$  & $<$ 19.9      & $<$ 23          & 400  & $-$1.95 & $-$3.55 \\
\ion{Na}{viii}$^{*}$ & $<$ 20.4       & $<$ 19.3 & 500  & 0.7    & $-$0.90 & \ion{Fe}{iv}$^{*}$   & $<$ 19.5      & $<$ 19.1        & 400  & $-$1.60 & $-$3.20 \\ 
\ion{Na}{ix}$^{*}$   & $<$ 20.1       & $<$ 23         & 600 & 1.25    & $-$0.35 & \ion{Fe}{v}$^{*}$    & $<$ 20.4      & $<$ 19.9        & 400  & $-$1.30 & $-$2.90 \\
\ion{Na}{x}$^{*}$    & $<$ 20.2       & 20.1 $\pm$ 0.3 & 700 & 1.7     & 0.10    & \ion{Fe}{vi}$^{*}$   & $<$ 23        & $<$ 19.4        & 400  & $-$1.00 & $-$2.60 \\
\ion{Na}{xi}$^{*}$   & $<$ 20.3       & $<$ 23         & 700 & 2.3     & 0.70    & \ion{Fe}{vii}$^{*}$  & 19.9 $\pm$ 0.6& $<$ 19.8        & 500  & $-$0.50 & $-$2.10 \\
\ion{Mg}{vi}$^{*}$   & $<$ 20.1       & $<$ 19         & 500 & $-$0.3  & $-$1.90 & \ion{Fe}{viii}$^{*}$ & 20.0 $\pm$ 0.3& $<$ 19.8        & 500  & $-$0.10 & $-$1.70 \\
\ion{Mg}{vii}$^{*}$  & 20.6 $\pm$ 0.9 & $<$ 19.7       & 500 & 0.1     & $-$1.50 & \ion{Fe}{ix}   & 20.2 $\pm$ 0.4& 20.97 $\pm$ 0.08& 500  & 0.20    & $-$1.40 \\
\ion{Mg}{viii}$^{*}$ & $<$ 20.2       & $<$ 19.8       & 500 & 0.5     & $-$1.10 & \ion{Fe}{x}    & 20.4 $\pm$ 0.1& 20.2 $\pm$ 0.1  & 500  & 0.40    & $-$1.20 \\ 
\ion{Mg}{ix}   & 20.3 $\pm$ 0.2 & 20.3 $\pm$ 0.3 & 560 & 1.0     & $-$0.60 & \ion{Fe}{xi}   & 20.1 $\pm$ 0.2& 19.8 $\pm$ 0.2  & 500  & 0.60    & $-$1.00 \\ 
\ion{Mg}{x}$^{*}$    & $<$ 20.2       & $<$ 19.4       & 700 & 1.4     & $-$0.20 & \ion{Fe}{xii}$^{*}$  & 19.7 $\pm$ 0.5& $<$ 19.7        & 600  & 0.80    & $-$0.80 \\
\ion{Mg}{xi}   & 20.9 $\pm$ 0.2 & 20.4 $\pm$ 0.2 & 500 & 1.9     & 0.30    & \ion{Fe}{xiii} & 19.5 $\pm$ 0.6& 19.4 $\pm$ 0.4  & 600   & 0.95    & $-$0.65 \\ 
\ion{Mg}{xii}  & 21.0 $\pm$ 0.1 & 20.8 $\pm$ 0.1 & 680 & 2.35    & 0.75    & \ion{Fe}{xiv}$^{*}$  & $<$ 19.7      & 19.3 $\pm$ 0.9  & 700  & 1.15    & $-$0.45 \\
\ion{Al}{vii}$^{*}$  & $<$ 20.8       & $<$ 19.9 & 400 & 0.05    & $-$1.55 & \ion{Fe}{xv}$^{*}$   & $<$ 19.6      & $<$ 19.0        & 700  & 1.40    & $-$0.20  \\
\ion{Al}{viii}$^{*}$ & 21.1 $\pm$ 0.2 & 20.7 $\pm$ 0.4 & 400 & 0.45& $-$1.15 & \ion{Fe}{xvi}$^{*}$  & $<$ 18.9      & $<$ 18.6        & 700  & 1.40    & $-$0.20   \\
\ion{Al}{ix}$^{*}$   & $<$ 19.6       & $<$ 19.3       & 600 & 0.9     & $-$0.70 & \ion{Fe}{xvii} & 20.7 $\pm$ 0.2& 20.5 $\pm$ 0.2  & 740  & 1.80    & 0.20    \\
\ion{Al}{xi}$^{*}$   & $<$ 19.3       & $<$ 19.5       & 700 & 1.7     & 0.10    & \ion{Fe}{xviii}& 20.3 $\pm$ 0.2& 20.1 $\pm$ 0.2  & 700  & 2.05    & 0.45    \\
\ion{Al}{xii}$^{*}$  & $<$ 20.0       & $<$ 19.9       & 700 & 2.05    & 0.45    & \ion{Fe}{xix}  & 20.7 $\pm$ 0.1& 20.5 $\pm$ 0.1  & 800  & 2.25    & 0.65    \\
\ion{Al}{xiii}$^{*}$ & $<$ 20.3       & $<$ 20.7       & 800 & 2.45    & 0.85    & \ion{Fe}{xx}   & 20.4 $\pm$ 0.2& 20.4 $\pm$ 0.1  & 800  & 2.40    & 0.80     \\
\ion{Si}{vii}$^{*}$  & 19.7 $\pm$ 1.0 & $<$ 19.8       & 400 & 0.05    & $-$1.55 & \ion{Fe}{xxi}  & 20.7 $\pm$ 0.2& 20.5 $\pm$ 0.2  & 800  & 2.60    & 1.00     \\
\ion{Si}{viii}$^{*}$ & 19.7 $\pm$ 0.3 & 20.9 $\pm$ 0.1 & 620 & 0.4     & $-$1.20 & \ion{Fe}{xxii} & 20.2 $\pm$ 0.2& 20.3 $\pm$ 0.3  & 800  & 2.75    & 1.15    \\
\ion{Si}{ix}$^{*}$   & 20.0 $\pm$ 0.4 & 21.1 $\pm$ 0.1 & 400 & 0.75    & $-$0.85 & \ion{Fe}{xxiii}& 20.5 $\pm$ 0.2& 20.4 $\pm$ 0.2  & 800  & 2.90    & 1.30    \\
\ion{Si}{x}    & 19.9 $\pm$ 0.2 & 20.0 $\pm$ 0.2 & 790 & 1.05    & $-$0.55 & \ion{Fe}{xxiv} & 20.7 $\pm$ 0.2& 20.8 $\pm$ 0.2  & 800  & 3.10    & 1.50      \\
\ion{Si}{xi}   & 19.7 $\pm$ 0.2 & 19.8 $\pm$ 0.4 & 810 & 1.5     & $-$0.10 & \ion{Ni}{i}$^{*}$    & 20.3 $\pm$ 0.2& 20.4 $\pm$ 0.2  & 400  & $< -$4& $< -$5.6    \\
\ion{Si}{xii}  & 20.6 $\pm$ 0.5 & 20.3 $\pm$ 0.5 & 800 & 1.75    & 0.15    & \ion{Ni}{ii}$^{*}$   & 20.0 $\pm$ 0.3& 20.0 $\pm$ 0.3  & 400  & $< -$4& $< -$5.6 \\ 
\ion{Si}{xiii} & 20.7 $\pm$ 0.1 & 20.5 $\pm$ 0.2 & 660 & 2.2     & 0.60    & \ion{Ni}{iii}$^{*}$  & 20.0 $\pm$ 0.7& $<$ 20.2        & 400  & $-$1.70 & $-$3.30 \\
\ion{Si}{xiv}  & 20.9 $\pm$ 0.1 & 20.7 $\pm$ 0.2 & 880 & 2.6     & 1.00    & \ion{Ni}{iv}$^{*}$   & $<$ 20.2      & $<$ 20.1        & 400  & $-$1.50 & $-$3.10 \\
\ion{S}{vii}$^{*}$   & $<$ 19.4       & $<$ 19.3       & 400 & $-$0.2  & $-$1.80 & \ion{Ni}{v}$^{*}$    & $<$ 19.9      & $<$ 19.9        & 400  & $-$1.40 & $-$3.00 \\
\ion{S}{viii}  & 19.8 $\pm$ 0.3 & 19.7 $\pm$ 0.2 & 400& 0.2     & $-$1.40 & \ion{Ni}{vi}$^{*}$   & $<$ 20.0      & 20.1 $\pm$ 0.2  & 400  & $-$1.20 & $-$2.80 \\
\ion{S}{ix}$^{*}$    & 19.9 $\pm$ 0.1 & 19.6 $\pm$ 0.1 & 600 & 0.6     & $-$1.00 & \ion{Ni}{vii}$^{*}$  & 20.3 $\pm$ 0.1& 20.3 $\pm$ 0.1  & 400  & $-$0.95 & $-$2.55 \\ 
\ion{S}{x}$^{*}$     & 20.3 $\pm$ 0.3 & 20.1 $\pm$ 0.2 & 600 & 1.0     & $-$0.60 & \ion{Ni}{viii}$^{*}$ & 19.9 $\pm$ 0.3& 20.1 $\pm$ 0.2  & 500  & $-$0.50 & $-$2.10 \\
\ion{S}{xi}    & 19.7 $\pm$ 0.3 & 19.7 $\pm$ 0.4 & 560 & 1.2     & $-$0.30 & \ion{Ni}{ix}$^{*}$   & 20.2 $\pm$ 0.1& 20.4 $\pm$ 0.1  & 500  & $-$0.05 & $-$1.65 \\
\ion{S}{xii}   & 19.7 $\pm$ 0.3 & 19.8 $\pm$ 0.4 & 620 & 1.45    & $-$0.15 & \ion{Ni}{x}$^{*}$    & $<$ 19.8      & 20.0 $\pm$ 0.3  & 500  & 0.40    & $-$1.20 \\
\ion{S}{xiii}  & 19.9 $\pm$ 0.9 & 19.7 $\pm$ 0.6 & 700 & 1.8     & 0.20    & \ion{Ni}{xi}$^{*}$   & 20.1 $\pm$ 0.4& 20.8 $\pm$ 0.1  & 600  & 0.70    & $-$0.90\\
\ion{S}{xiv}$^{*}$   & 20.3 $\pm$ 0.3 & $<$ 20.0       & 700 & 2.15    & 0.55    & \ion{Ni}{xii}$^{*}$  & 20.3 $\pm$ 0.1& 20.4 $\pm$ 0.1  & 600  & 0.90    & $-$0.70 \\ 
\ion{S}{xv}$^{*}$    & $<$ 20.7       & $<$ 20.1       & 800 & 2.4     & 0.80    & \ion{Ni}{xiii}$^{*}$ & 20.0 $\pm$ 0.2& 20.0 $\pm$ 0.2  & 600  & 1.10    & $-$0.50 \\
\ion{S}{xvi}$^{*}$   & $<$ 20.8       & 21.0 $\pm$ 0.4 & 800 & 2.85    & 1.25    & \ion{Ni}{xiv}$^{*}$ & $<$ 19.7      & 19.5 $\pm$ 0.5  & 700  & 1.25    & $-$0.25 \\
\ion{Ar}{ix}$^{*}$   & 19.5 $\pm$ 0.5 & 19.6 $\pm$ 0.3 & 500 & 0.4& $-$1.20 & \ion{Ni}{xv}$^{*}$   & $<$ 18.9      & $<$ 19.3        & 700  & 1.50    & $-$0.10 \\
\ion{Ar}{x}$^{*}$    & 19.9 $\pm$ 0.3 & 19.4 $\pm$ 0.3 & 500 & 0.8     & $-$0.75 & \ion{Ni}{xvi}$^{*}$  & $<$ 19.4      & 19.7 $\pm$ 0.2  & 700  & 1.40    & $-$0.20 \\
\ion{Ar}{xi}$^{*}$   & 19.9 $\pm$ 0.2 & 19.7 $\pm$ 0.2 & 600 & 1.1    & $-$0.40 & \ion{Ni}{xvii}$^{*}$ & 19.6 $\pm$ 0.4& 20.8 $\pm$ 0.3  & 700  & 1.60    & 0.00    \\
\ion{Ar}{xii}$^{*}$  & 19.3 $\pm$ 0.8 & 19.2 $\pm$ 0.4 & 600 & 1.3     & $-$0.20 & \ion{Ni}{xviii}$^{*}$& 19.7 $\pm$ 0.3& 19.7 $\pm$ 0.3  & 700  & 1.70    & 0.10   \\
\end{tabular}
}
\end{center}
\end{table*}

\newpage
\clearpage

\section*{APPENDIX B}
In this Appendix we discuss those ions for which the column density may be affected by the effects described in Sect.~\ref{sect:id}.
{\it Carbon, Oxygen and Neon:} For \ion{C}{iv} the strongest line at 41.42 \AA~coincides with the instrumental \ion{C}{i} edge, which is deep and susceptible to calibration uncertainties. For \ion{O}{iv} the wavelength of the dominant line in the X-ray range is not well known and the absorption line is weak.  For \ion{Ne}{viii} the deepest line at 88.10~\AA~is at a rather noisy part of the spectrum, the other line at 67.38 \AA~is blended with an \ion{Al}{viii} line.
\par  
{\it Sodium and Aluminum:} Both \ion{Na}{vii} and \ion{Na}{viii} are only detected through lines above 60 \AA. The column density for \ion{Al}{vii} is mainly determined from the 2p$_{1/2}$ edge at 43.56~\AA~and several lines above 60 \AA. A similar situation occurs for \ion{Al}{viii}, where the 2p$_{1/2}$ and 2p$_{3/2}$ edges at 51.36 \AA~dominate in the column density determination. 
\par
{\it Silicon and Sulfur:} The deepest lines for \ion{Si}{viii} and \ion{Si}{ix} at 61.04 \AA~and 55.30 \AA, respectively are saturated. For \ion{Si}{viii} this line is also blended. The saturation of these well detected lines explains the difference in column density obtained for the different velocity broadening assumed. For \ion{S}{x} the strongest lines form a blend in the instrumental \ion{C}{i} edge. 
\par
{\it Argon:} The column density of \ion{Ar}{ix} is determined from an absorption line which coincides with the instrumental \ion{C}{i} edge and the \ion{C}{v} forbidden line. \ion{Ar}{x} has only one detectable line at 37.45 \AA, which is at a rather noisy part of the spectrum. For \ion{Ar}{xi} the deepest line is blended with the $-$1040~km~s$^{-1}$ component of \ion{S}{x} at 34.29 \AA. \ion{Ar}{xii} has only one weak line in the spectrum at 31.37 \AA, and for \ion{Ar}{xiii} we detect only two weak lines at 29.27 \AA~and 27.44 \AA. The deepest \ion{Ar}{xv} absorption line is blended with the \ion{N}{vii} Ly$\alpha$ line. The blending is further complicated by the possible substructure in the \ion{N}{vii} Ly$\alpha$ line due to the different outflow velocities. \ion{Ar}{xvi} is detected from only two edges in the spectrum at 15 and 25 \AA. 
\par
{\it Calcium:} \ion{Ca}{xiv} has only one unblended line at 21.13 \AA. However, the fit is driven by only one data point of the LETGS spectrum, making the column density uncertain. For all the other calcium ions the detection is very sensitive to the velocity broadening. If we leave $\sigma$ free, we mostly find upper limits to the velocity broadening of 30~km~s$^{-1}$, implying highly saturated lines and therefore unrealistically high column densities. The lower ionized states of calcium (\ion{Ca}{i} to \ion{Ca}{viii}) have edges in the spectra, which dominate the fit.
\par
{\it Nickel:} The detection of all nickel ions is uncertain. Nickel only produces weak absorption lines, most of them are blended with stronger iron absorption lines. Small errors in wavelength for these iron lines could be compensated in our model by absorption from nickel. For \ion{Ni}{i} the column density is determined from the observed short wavelength tail of the edge at 105.97 \AA~and a smaller edge at 14.42 \AA. For \ion{Ni}{ii} the 3s edge at 87.93 \AA~dominates the column density determination. The strongest \ion{Ni}{vi} and \ion{Ni}{vii} lines are blended by \ion{Fe}{xvi} at 14.30~\AA~and \ion{Fe}{xviii} at 14.26~\AA, respectively. The strongest \ion{Ni}{ix} line is blended by an \ion{Fe}{xx} line at 13.73 \AA. The column density for \ion{Ni}{xi} was determined mainly from the 3p-edge at 38.62~\AA, close to the instrumental \ion{C}{i} edge, and in a noisy part of the spectrum. The two strongest lines of \ion{Ni}{xii} are blended with the \ion{Ne}{ix} resonance line at 13.45 \AA. \ion{Ni}{xiii} causes a slight depression in the spectrum due to many weak absorption lines. However, the detection is uncertain as the continuum spectrum of the MEG around the relevant wavelength of 13.27 \AA~differs by about 5 \% from the LETGS spectrum, a similar level as the expected depression. The deepest line for \ion{Ni}{xvii} is blended with an \ion{Fe}{xx} at 12.90 \AA; \ion{Ni}{xviii} has several weaker lines that are blended with \ion{Fe}{xx} at 12.60 \AA. 
\par
Due to the uncertainties in the determined column densities for the above listed ions, we decided not to use them in the further analysis. 

\section*{\bf APPENDIX C}

Here we present the LETGS, MEG and HEG spectra of NGC~5548. The dotted line indicate the zero flux level for the MEG and HEG spectra. The HEG spectrum was rebinned by a factor of 2 for plotting purposes, the MEG data was rebinned by a factor of 2 above 19 \AA. The Fe K$\alpha$ line appears slightly shifted in wavelength between the HEG and the MEG spectra, due to calibration uncertainties or noise in the MEG spectrum. Between 4 \AA~and 6.2 \AA~there is a calibration mismatch between the HEG and the MEG spectra. The thick line is for the model with broadened emission lines, the thin line is the same model but without broadened emission lines. 

\begin{figure*}
\begin{tabular}{lr}
\parbox{9truecm}{\includegraphics[width=6.5cm,height=8cm,angle=-90]{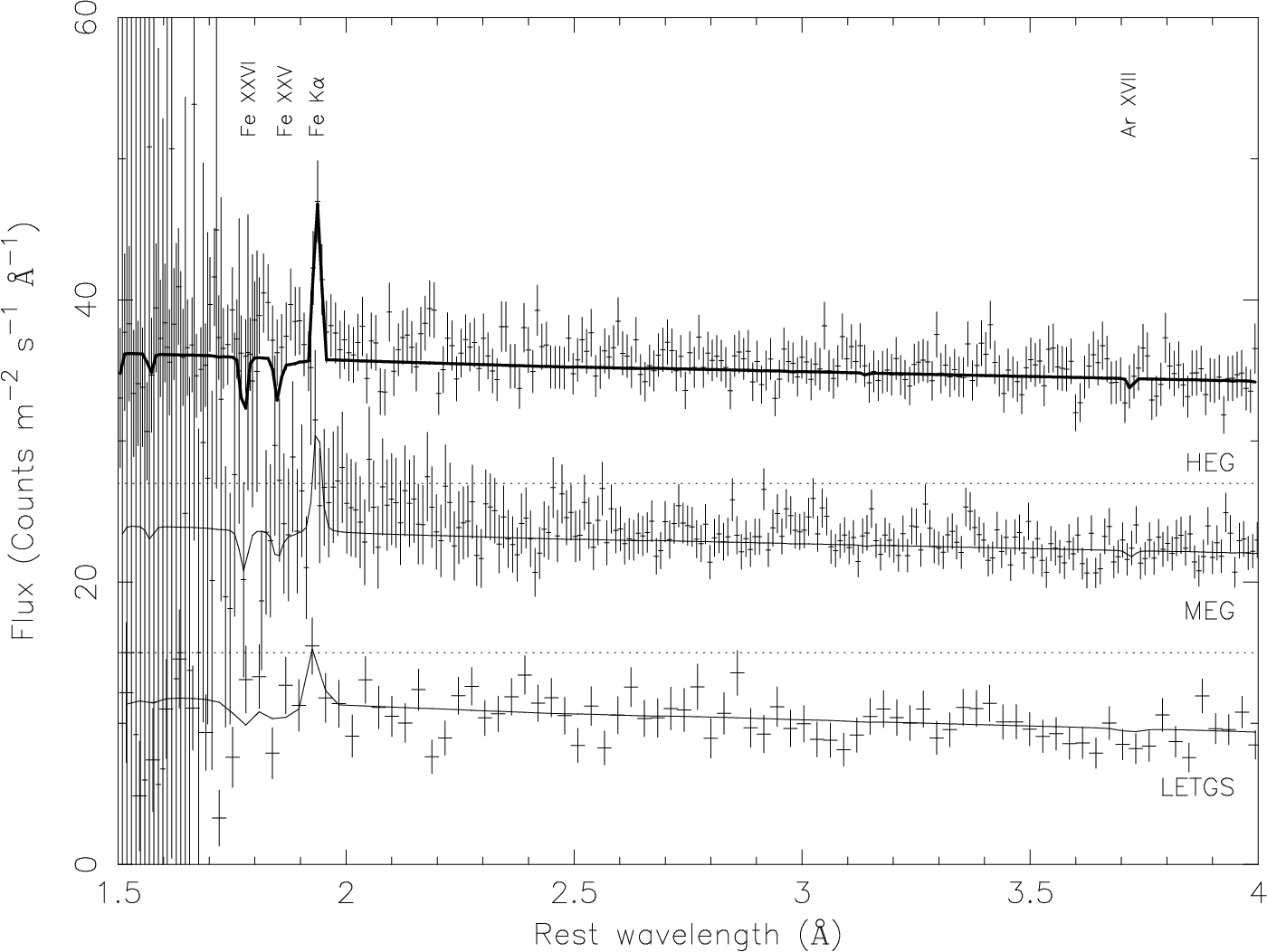}} & \parbox{9truecm}{\includegraphics[width=6.5cm,height=8cm,angle=-90]{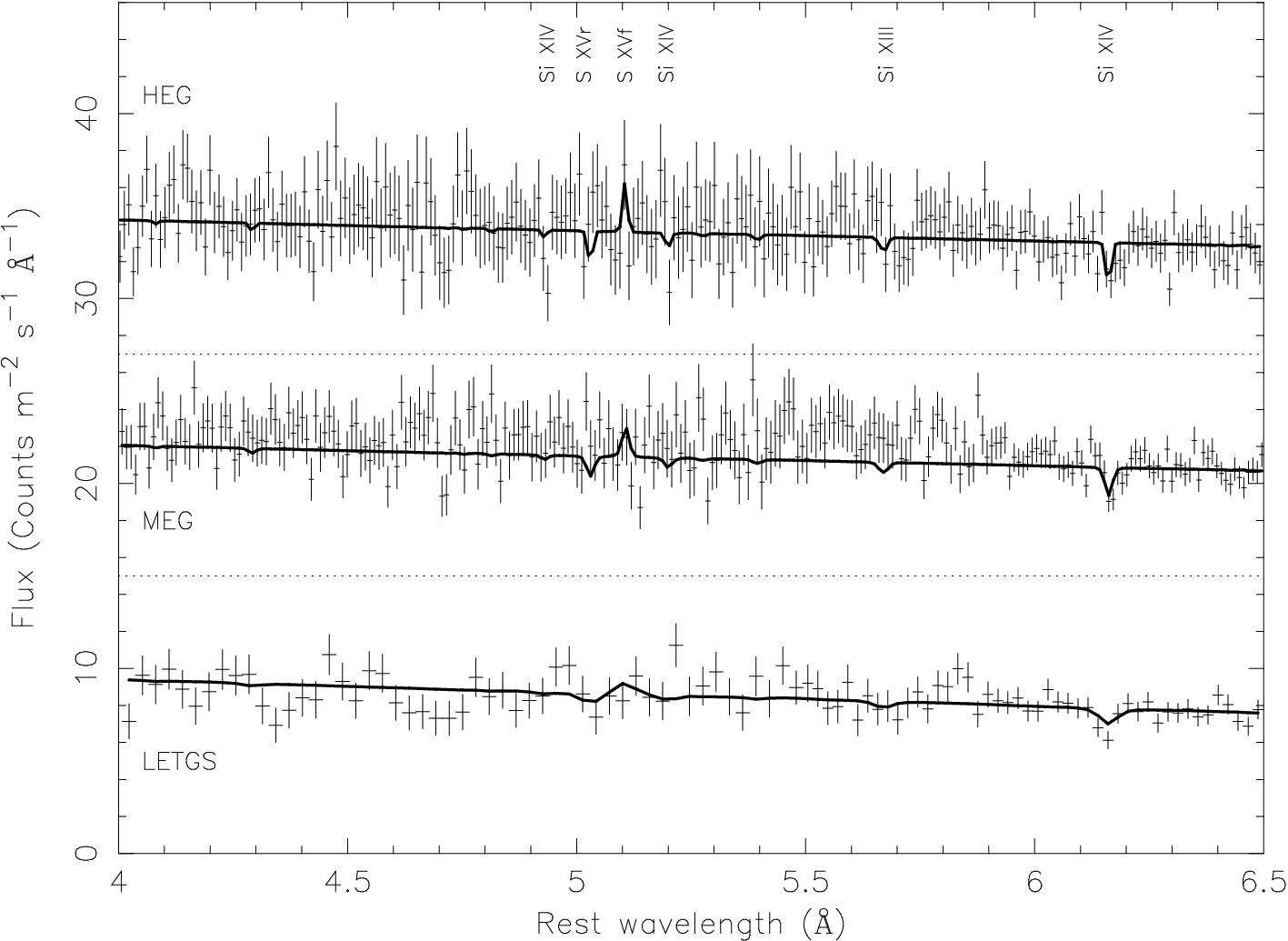}} \\
\ \hspace{1truecm} &  \\
\parbox{9truecm}{\includegraphics[width=6.5cm,height=8cm,angle=-90]{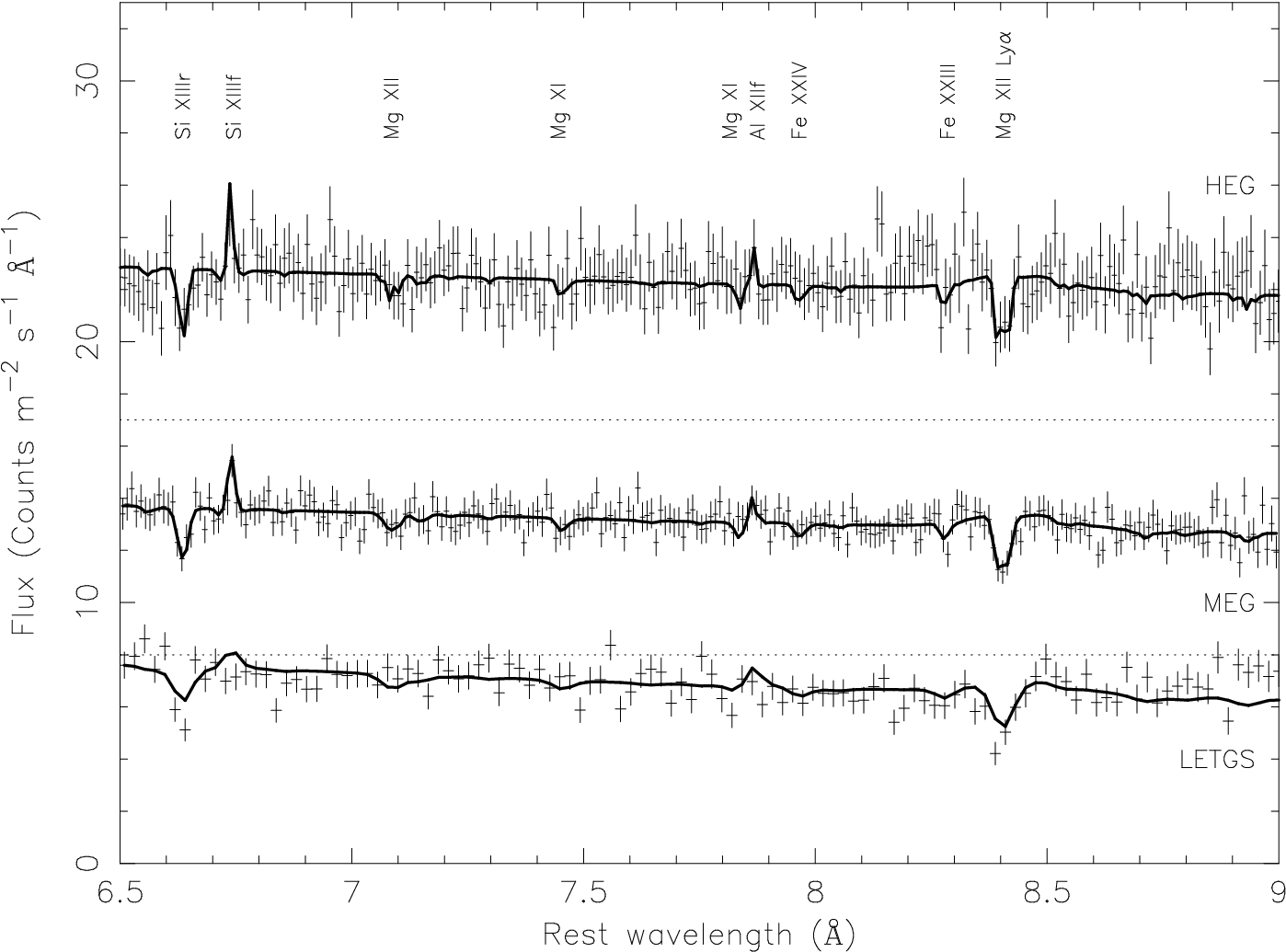}} & \parbox{9truecm}{\includegraphics[width=6.5cm,height=8cm,angle=-90]{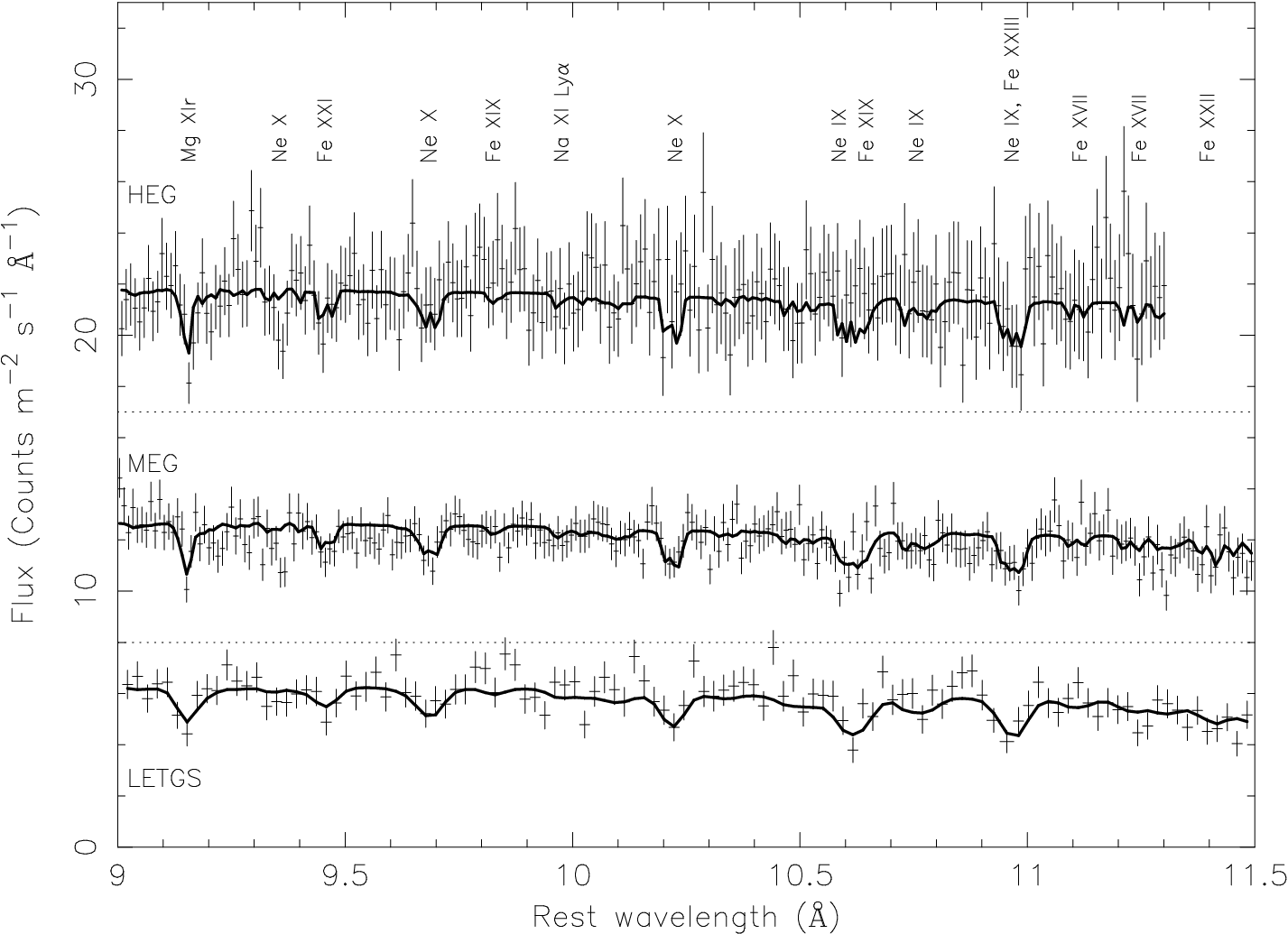}} \\
\ \hspace{1truecm} &  \\
\parbox{9truecm}{\includegraphics[width=6.5cm,height=8cm,angle=-90]{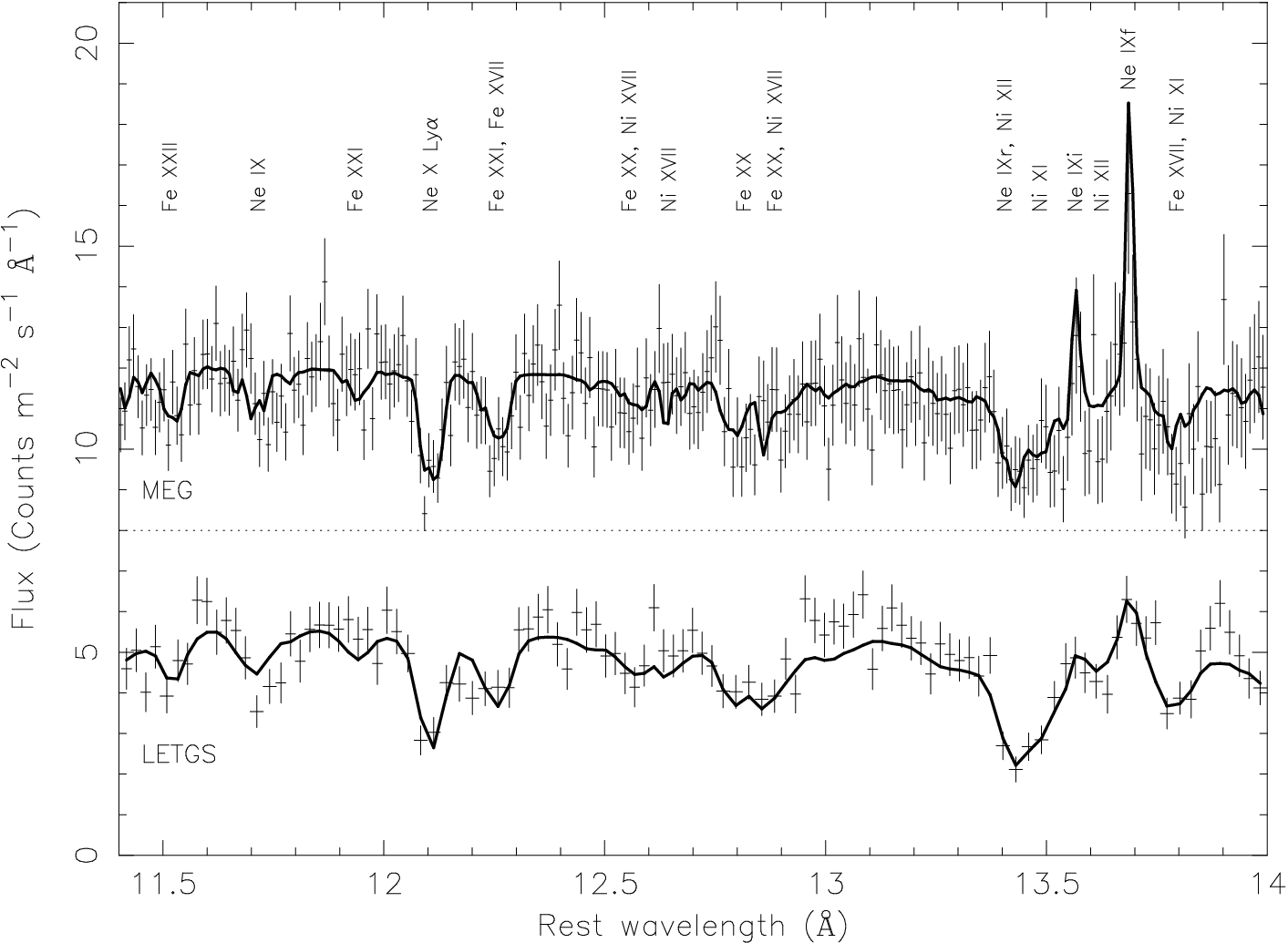}} & \parbox{9truecm}{\includegraphics[width=6.5cm,height=8cm,angle=-90]{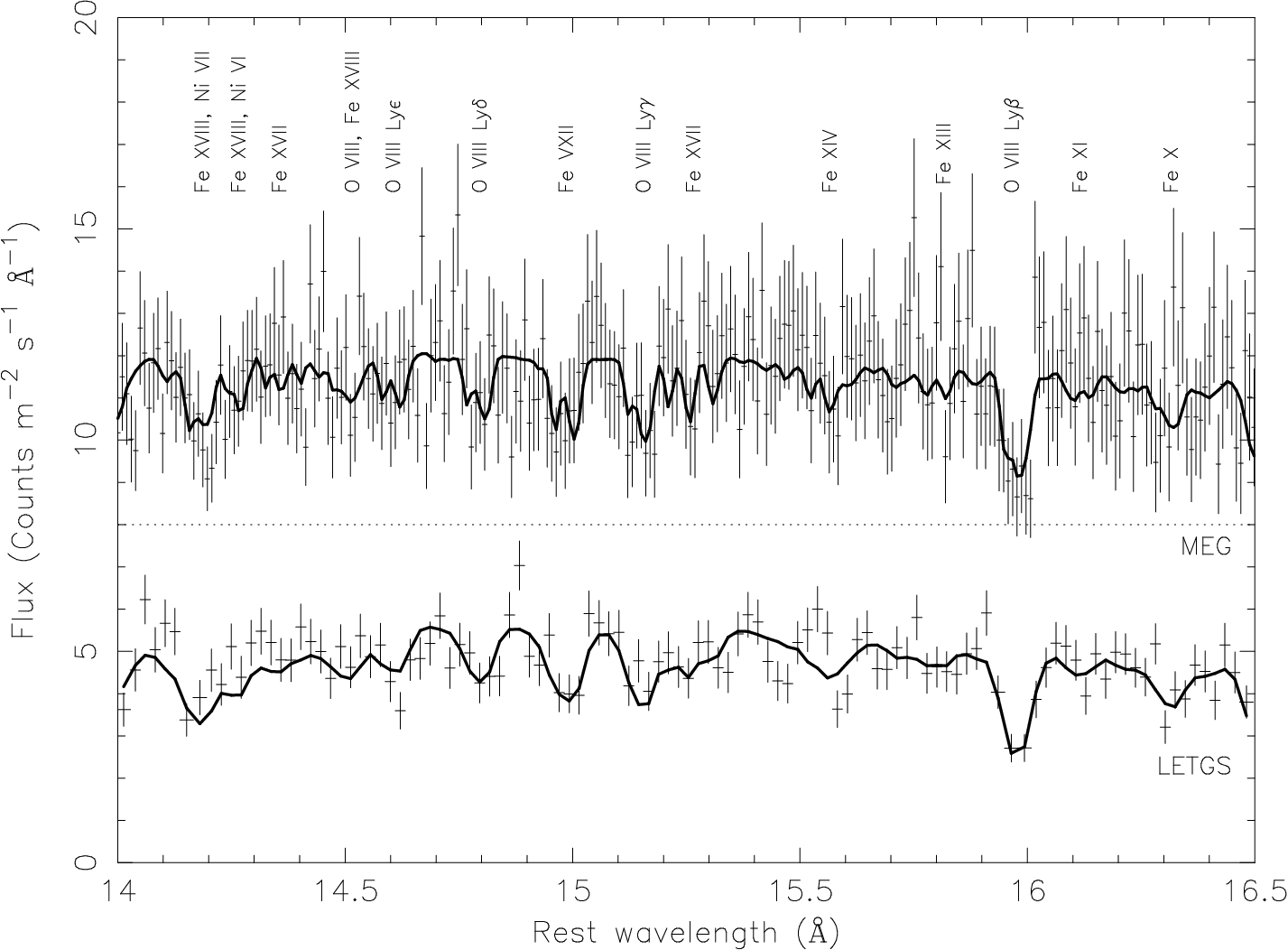}} \\
\end{tabular}
\caption{Detail of the HEG, MEG and LETGS spectra. The dotted lines indicate the zero flux level for the HEG and MEG spectra.}
\label{fig:shift1}
\end{figure*}

\newpage
\addtocounter{figure}{-1}
\clearpage

\begin{figure*}
\begin{tabular}{lr}
\parbox{9truecm}{\includegraphics[width=6.5cm,height=8cm,angle=-90]{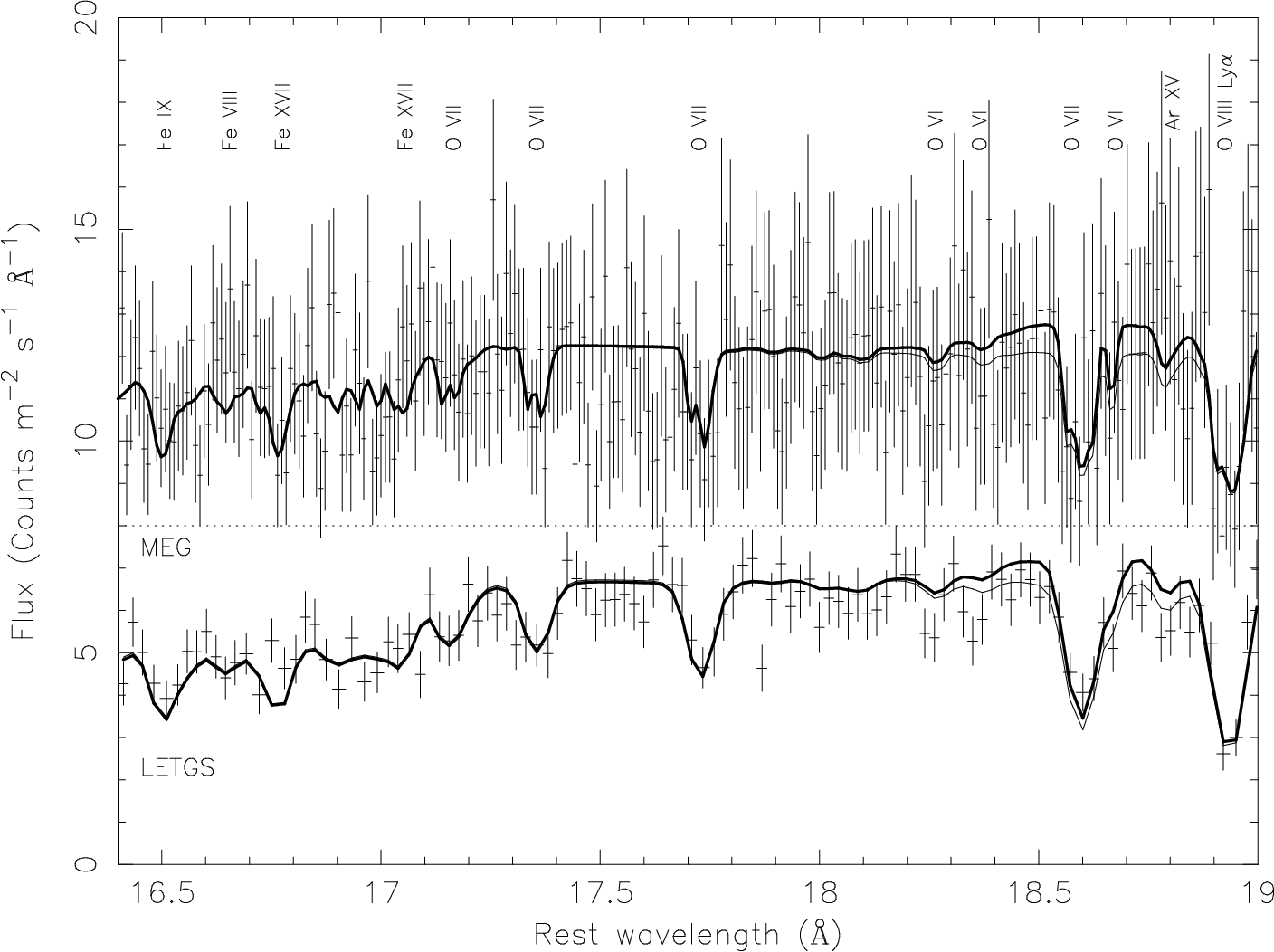}} & \parbox{9truecm}{\includegraphics[width=6.5cm,height=8cm,angle=-90]{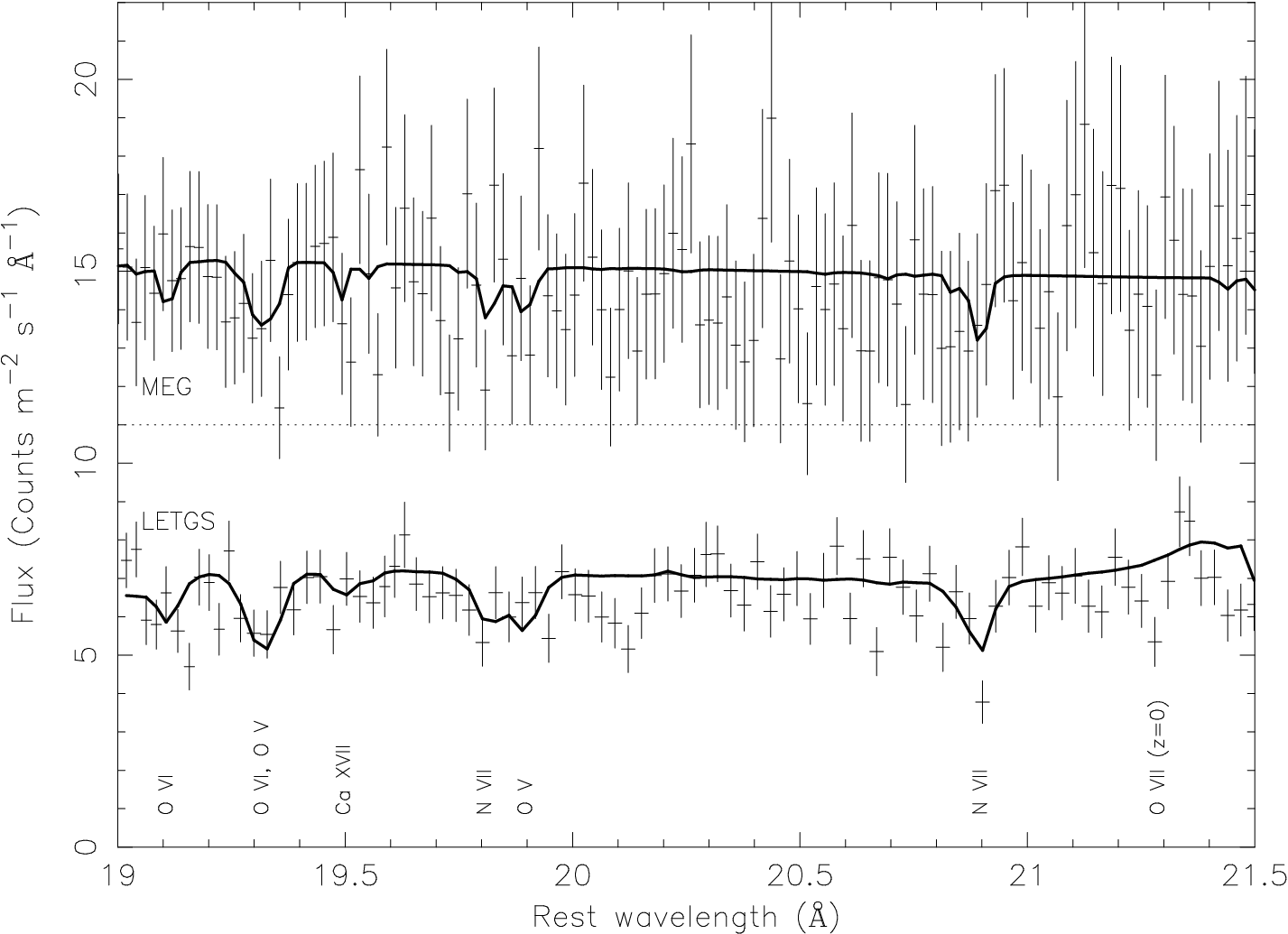}}  \\
\ \hspace{1truecm} &  \\
\parbox{9truecm}{\includegraphics[width=6.5cm,height=8cm,angle=-90]{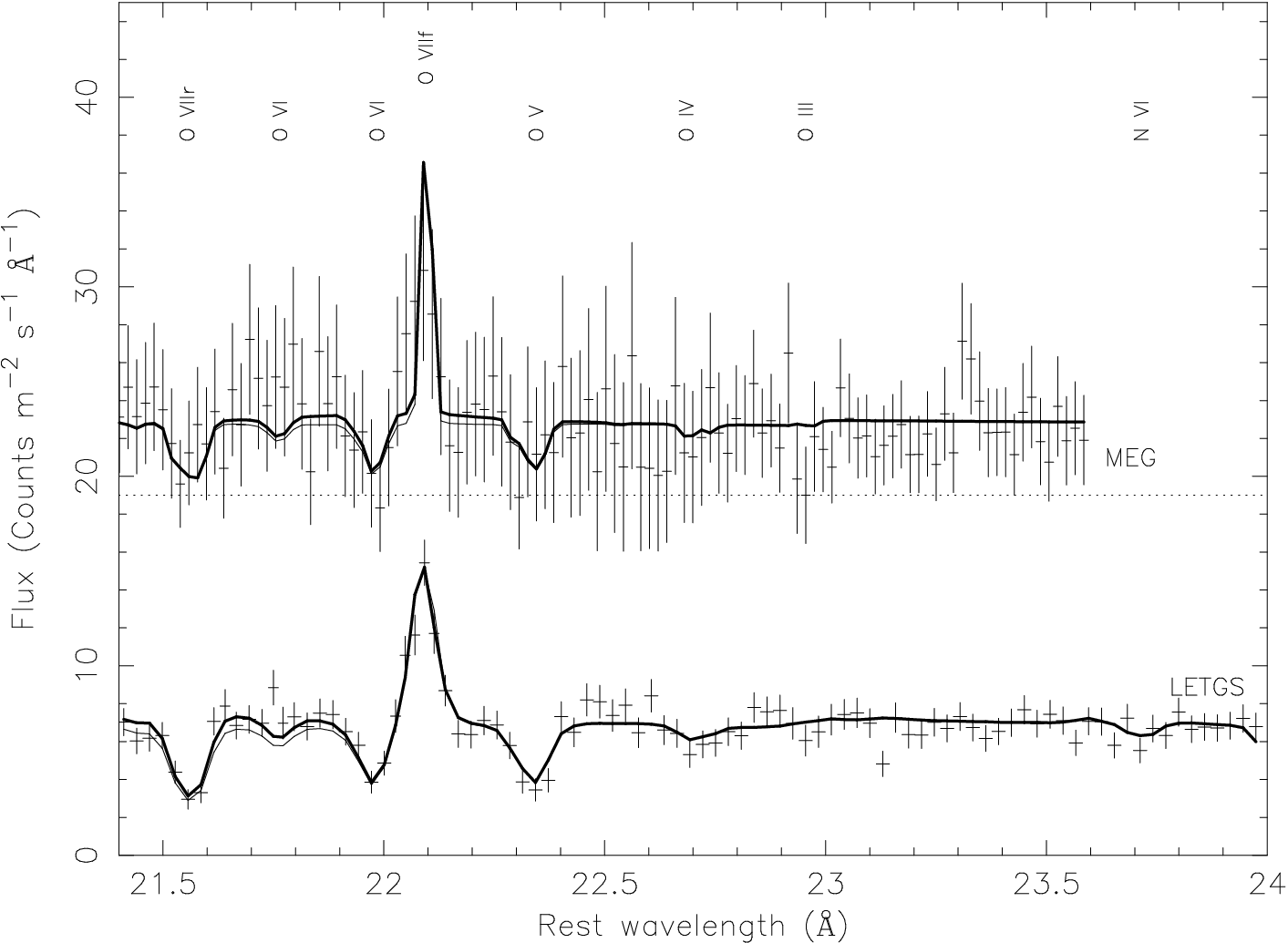}} & \parbox{9truecm}{\includegraphics[width=6.5cm,height=8cm,angle=-90]{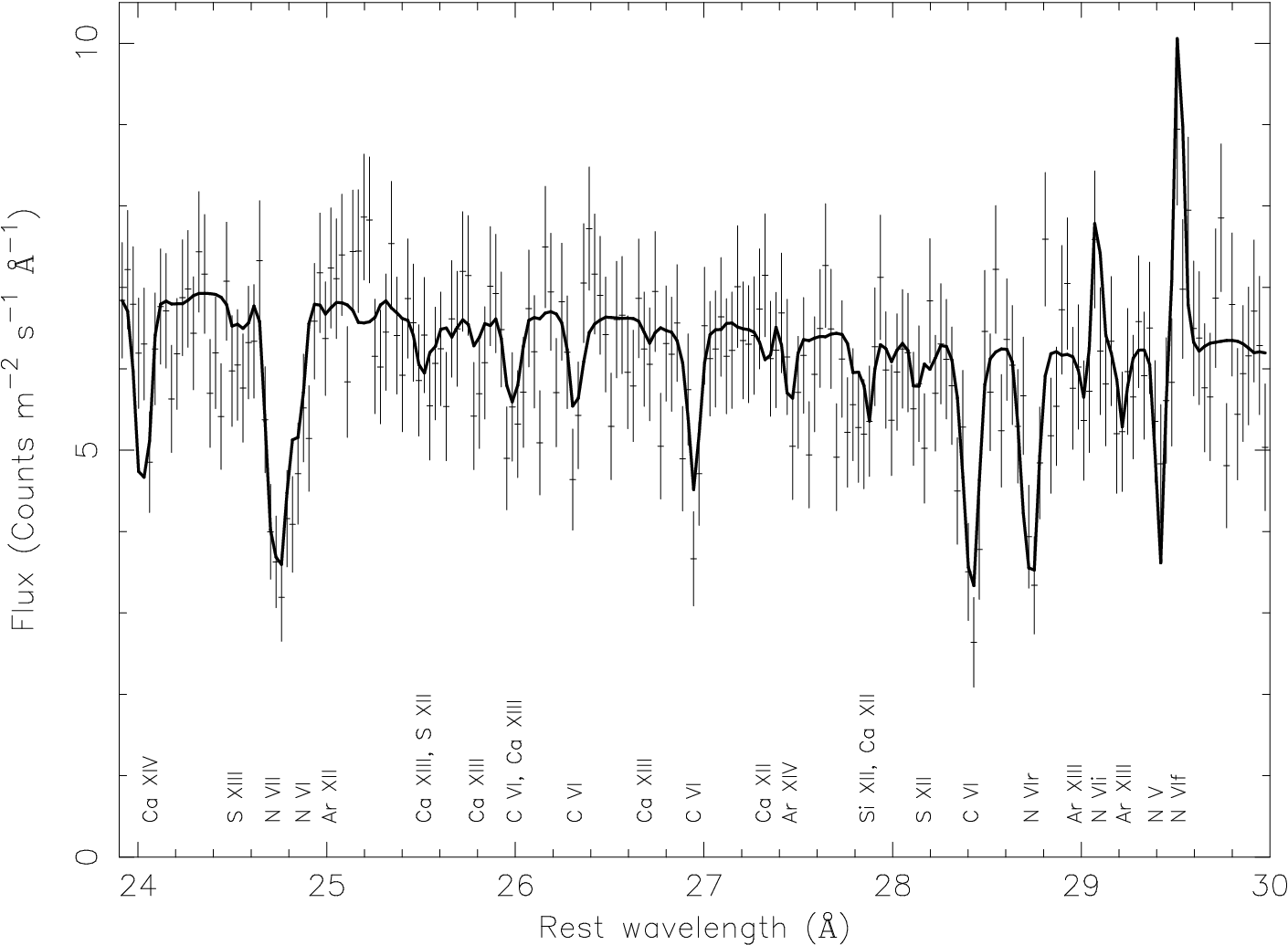}}  \\
\ \hspace{1truecm} &  \\
\parbox{9truecm}{\includegraphics[width=6.5cm,height=8cm,angle=-90]{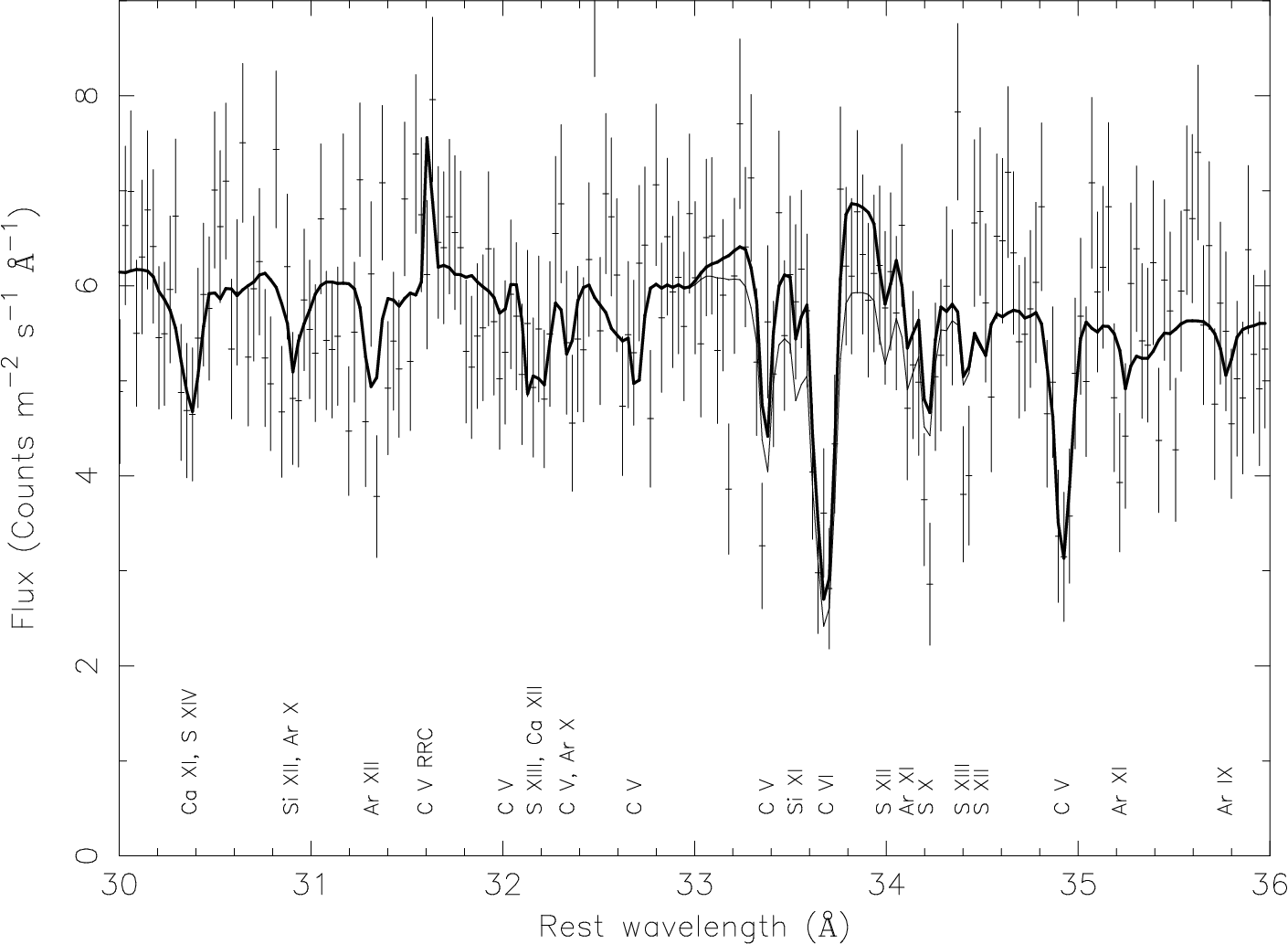}} & \parbox{9truecm}{\includegraphics[width=6.5cm,height=8cm,angle=-90]{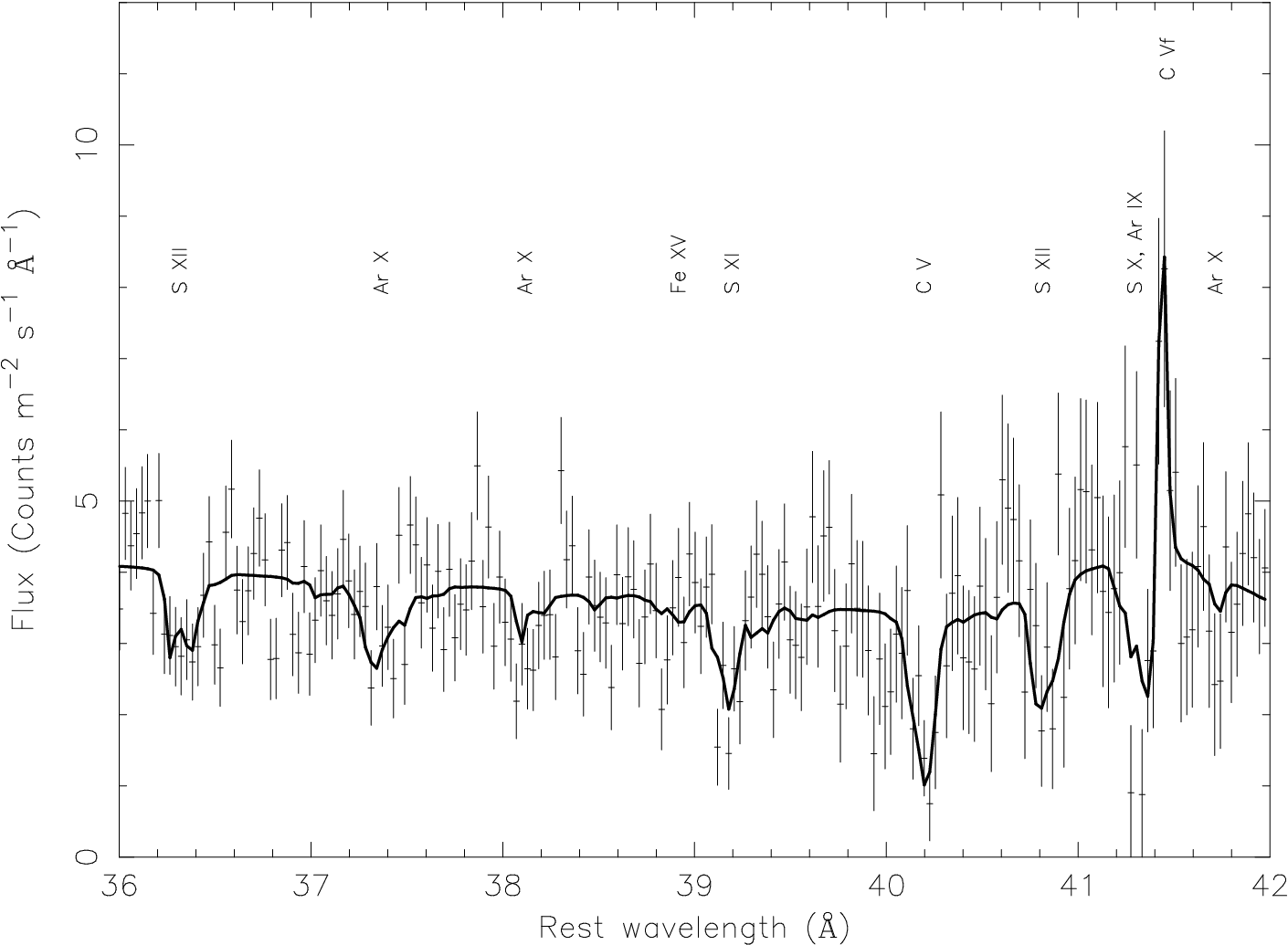}}  \\
\end{tabular}
\caption{Continued.}
\end{figure*}

\newpage
\addtocounter{figure}{-1}
\clearpage

\begin{figure*}
\begin{tabular}{lr}
\parbox{9truecm}{\includegraphics[width=6.5cm,height=8cm,angle=-90]{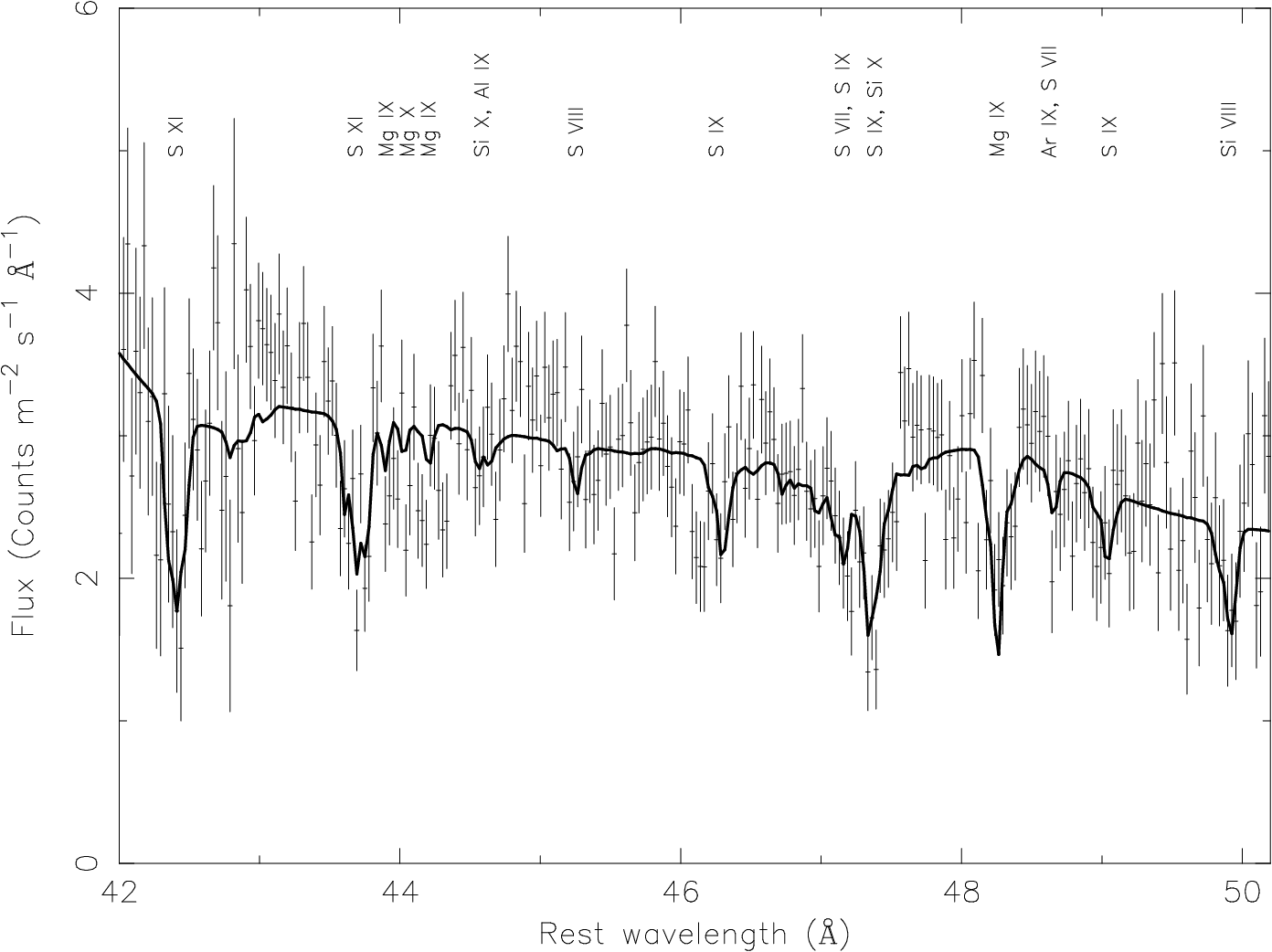}} & \parbox{9truecm}{\includegraphics[width=6.5cm,height=8cm,angle=-90]{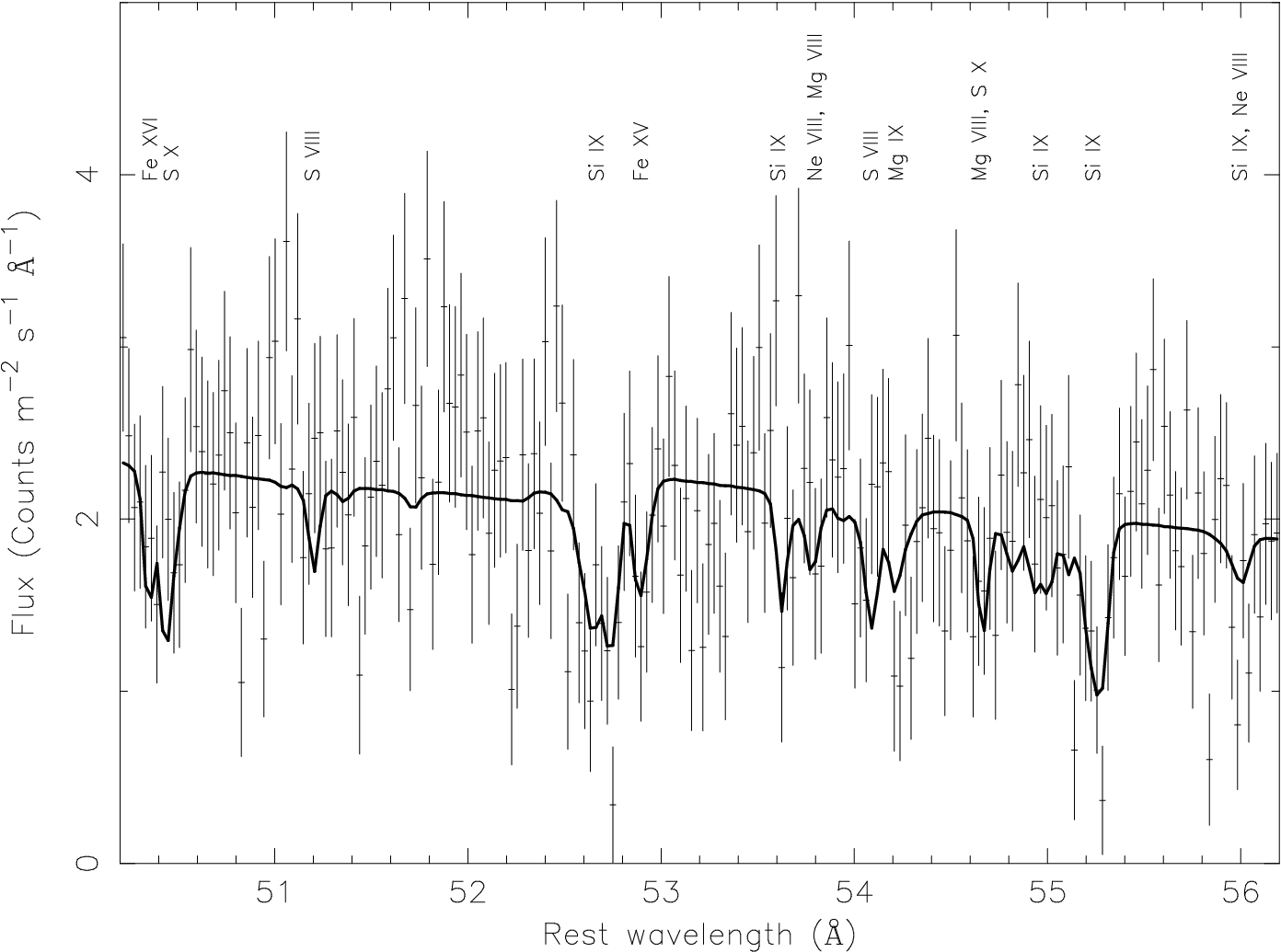}}  \\
\ \hspace{1truecm} &  \\
\parbox{9truecm}{\includegraphics[width=6.5cm,height=8cm,angle=-90]{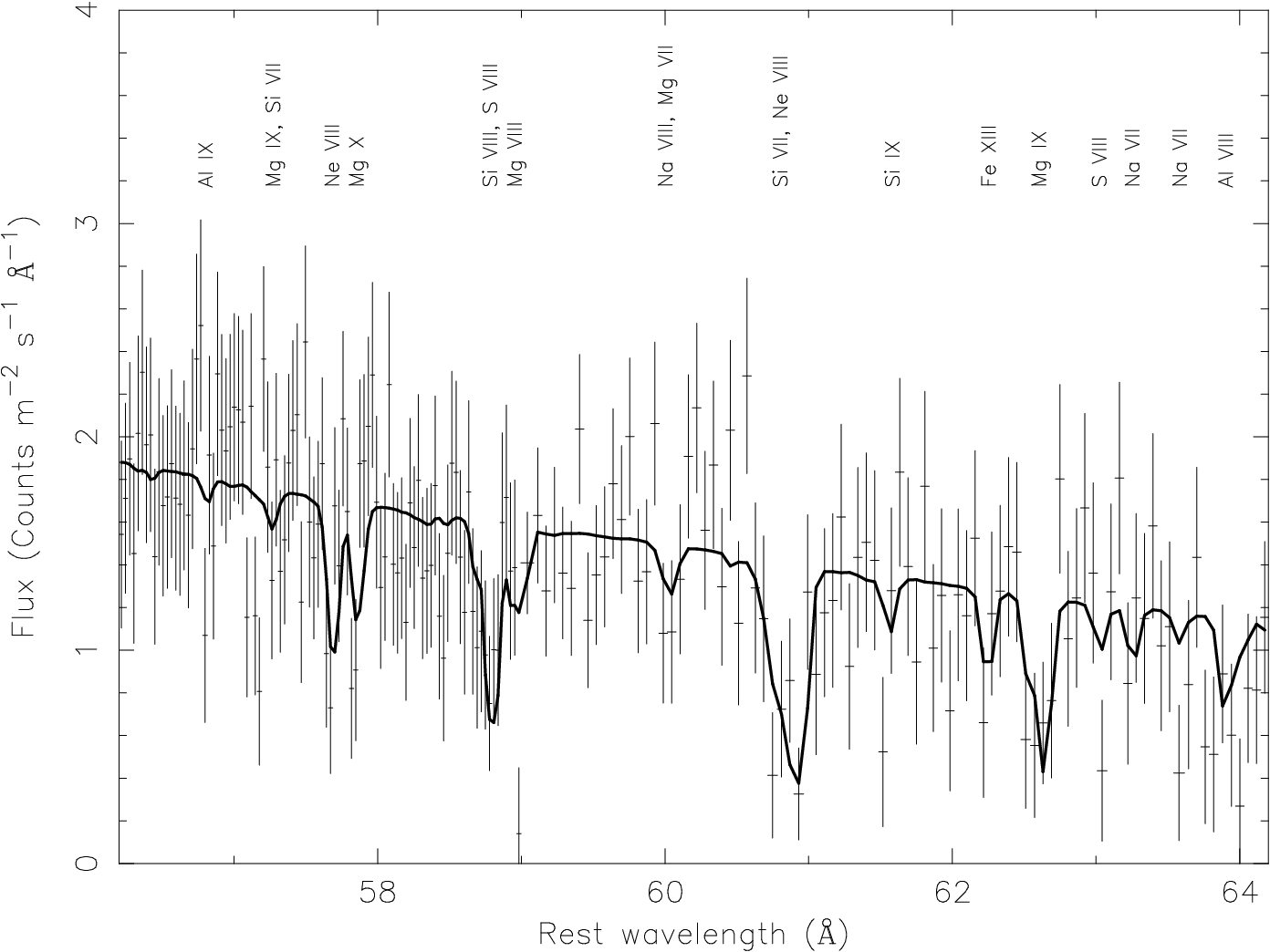}} & \parbox{9truecm}{\includegraphics[width=6.5cm,height=8cm,angle=-90]{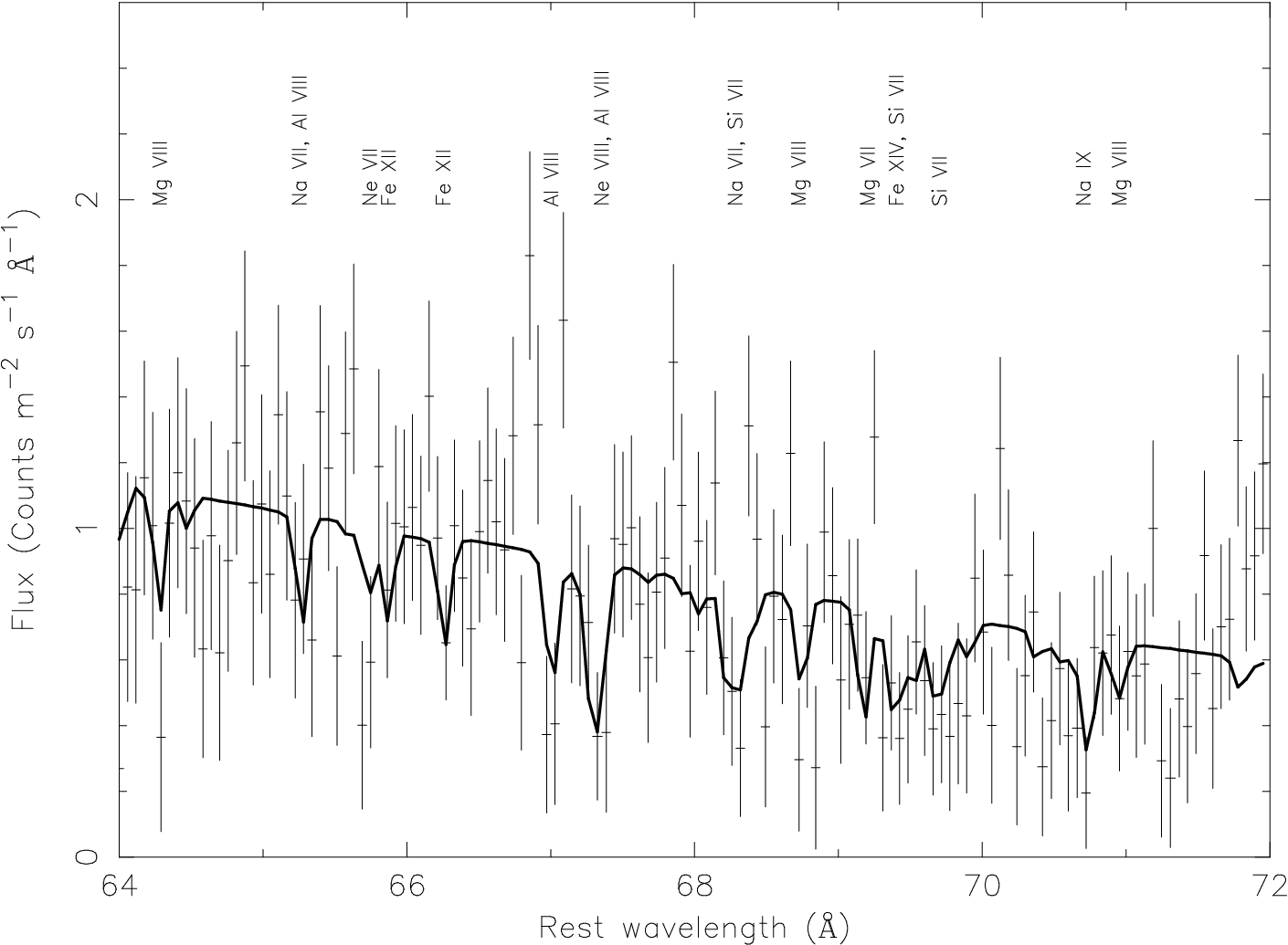}} \\
\ \hspace{1truecm} &  \\
\parbox{9truecm}{\includegraphics[width=6.5cm,height=8cm,angle=-90]{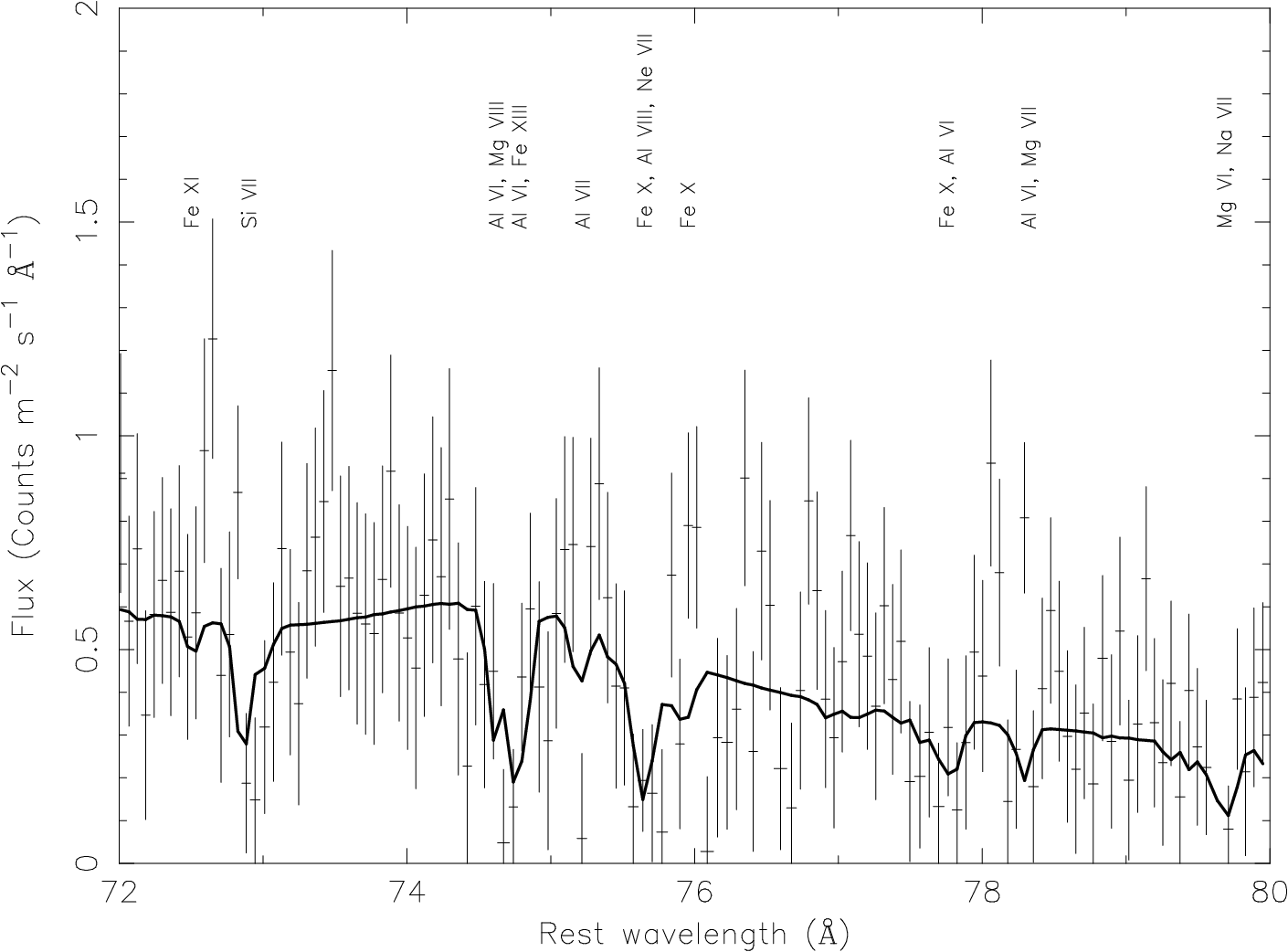}} & \parbox{9truecm}{\includegraphics[width=6.5cm,height=8cm,angle=-90]{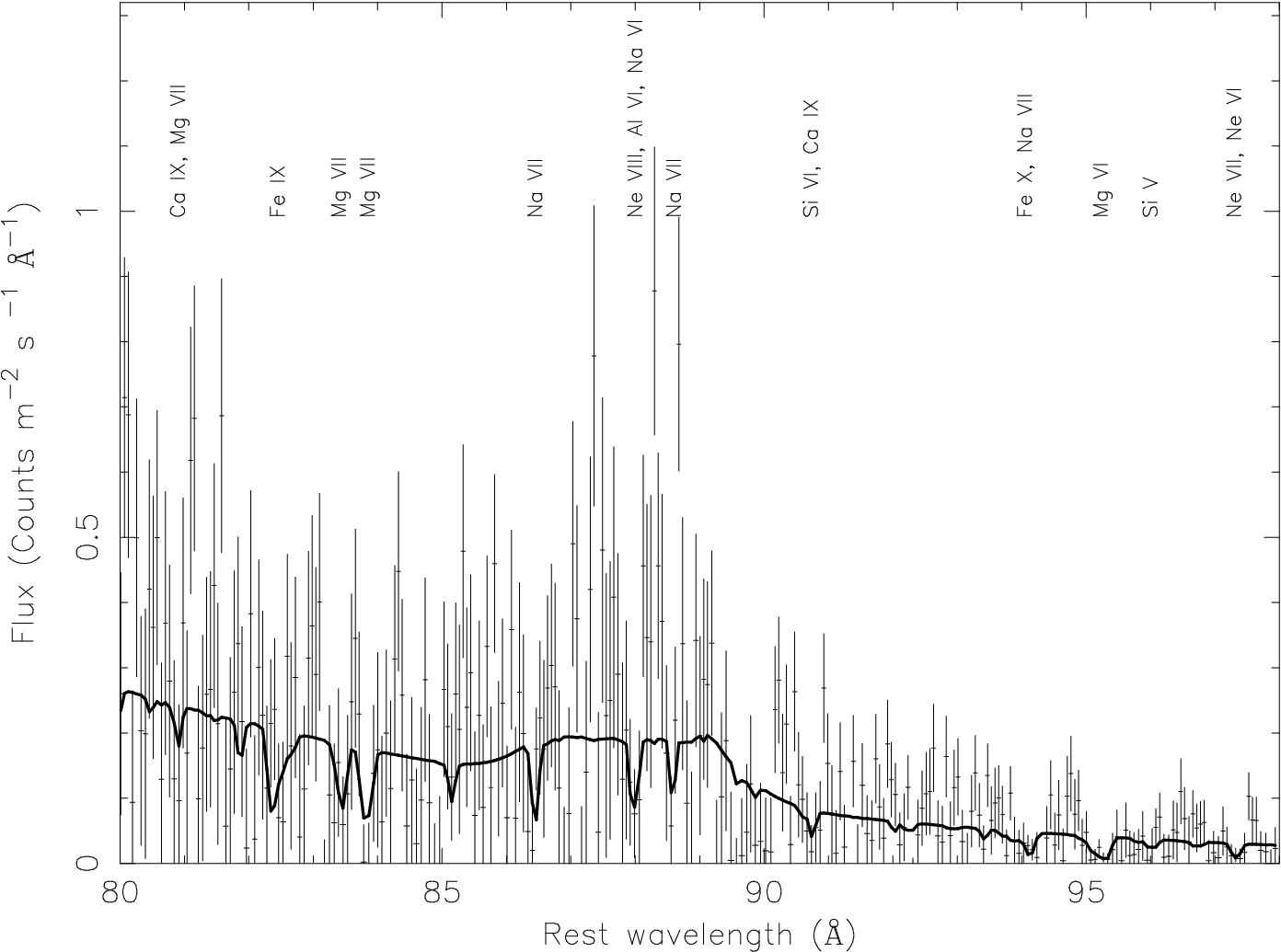}} \\
\end{tabular}
\caption{Continued.}
\label{fig:shift3}
\end{figure*}

\end{document}